\documentclass[12pt,english]{article}
\usepackage{lscape}
\usepackage[utf8]{inputenc}
\usepackage{graphicx}
\usepackage[titletoc,title]{appendix}
\usepackage{amsmath,amssymb,bm}
\usepackage[hidelinks]{hyperref}
\usepackage[export]{adjustbox}[2011/08/13]
\usepackage{chngcntr}
\usepackage{gensymb}

\usepackage[style=numeric,sorting=none,maxbibnames=99]{biblatex}

\allowdisplaybreaks
\usepackage{makecell}
\usepackage{xcolor}
\usepackage{upgreek}
\usepackage{siunitx}
\usepackage{lineno}
\usepackage{yfonts}

\DeclareMathOperator{\Var}{Var}
\DeclareMathOperator{\Cov}{Cov}

\usepackage{booktabs,threeparttable}
\usepackage{enumitem}

\usepackage{multirow}
\usepackage{booktabs}
\usepackage{url}
\usepackage{cancel}
\usepackage{soul} 
 
\usepackage{setspace}
\usepackage[hang]{footmisc}
\usepackage{algorithm}
\usepackage{algpseudocode}
\algrenewcommand\textproc{}

\renewcommand{\algorithmicensure}{\textbf{Output:}}
\usepackage{calc} 
\algrenewcommand\algorithmicensure{%
  \makebox[\widthof{\textbf{Require:}}][l]{\textbf{Ensure:}}}
\let\oldReturn\Return
\renewcommand{\Return}{\State\oldReturn}
\algdef{SE}{Begin}{End}{\textbf{begin}}{\textbf{end}}
\makeatletter
\newcommand*{\algrule}[1][\algorithmicindent]{\makebox[#1][l]{\hspace*{.5em}\thealgruleextra\vrule height \thealgruleheight depth \thealgruledepth}}%
\newcommand*{\thealgruleextra}{}
\newcommand*{\thealgruleheight}{.75\baselineskip}
\newcommand*{\thealgruledepth}{.25\baselineskip}

\newcount\ALG@printindent@tempcnta
\def\ALG@printindent{%
    \ifnum \theALG@nested>0
        \ifx\ALG@text\ALG@x@notext
        \else
            \unskip
            \addvspace{-1pt}
            \ALG@printindent@tempcnta=1
            \loop
                \algrule[\csname ALG@ind@\the\ALG@printindent@tempcnta\endcsname]%
                \advance \ALG@printindent@tempcnta 1
            \ifnum \ALG@printindent@tempcnta<\numexpr\theALG@nested+1\relax
            \repeat
        \fi
    \fi
    }%
\usepackage{etoolbox}
\patchcmd{\ALG@doentity}{\noindent\hskip\ALG@tlm}{\ALG@printindent}{}{\errmessage{failed to patch}}

\newbox\statebox
\newcommand{\myState}[1]{%
    \setbox\statebox=\vbox{#1}%
    \edef\thealgruleheight{\dimexpr \the\ht\statebox+1pt\relax}%
    \edef\thealgruledepth{\dimexpr \the\dp\statebox+1pt\relax}%
    \ifdim\thealgruleheight<.75\baselineskip
        \def\thealgruleheight{\dimexpr .75\baselineskip+1pt\relax}%
    \fi
    \ifdim\thealgruledepth<.25\baselineskip
        \def\thealgruledepth{\dimexpr .25\baselineskip+1pt\relax}%
    \fi
    \State #1%
    \def\thealgruleheight{\dimexpr .75\baselineskip+1pt\relax}%
    \def\thealgruledepth{\dimexpr .25\baselineskip+1pt\relax}%
}

\usepackage[a4paper, total={7in, 9in}]{geometry}
\usepackage[labelfont=bf]{caption}
\graphicspath{ {./images/} }

\usepackage{csquotes}
\usepackage[english]{babel}
\usepackage{comment}
\makeatletter
\def\@biblabel#1{}
\makeatother

\usepackage{wrapfig}
\usepackage{rotating}
\usepackage{epstopdf}

\usepackage{array}
\newcolumntype{L}[1]{>{\raggedright\let\newline\\\arraybackslash\hspace{0pt}}m{#1}}
\newcolumntype{C}[1]{>{\centering\let\newline\\\arraybackslash\hspace{0pt}}m{#1}}
\newcolumntype{R}[1]{>{\raggedleft\let\newline\\\arraybackslash\hspace{0pt}}m{#1}}

\newcommand*\patchAmsMathEnvironmentForLineno[1]{
  \expandafter\let\csname old#1\expandafter\endcsname\csname #1\endcsname
  \expandafter\let\csname oldend#1\expandafter\endcsname\csname end#1\endcsname
  \renewenvironment{#1}
  {\linenomath\csname old#1\endcsname}
  {\csname oldend#1\endcsname\endlinenomath}}
  \newcommand*\patchBothAmsMathEnvironmentsForLineno[1]{
  \patchAmsMathEnvironmentForLineno{#1}
  \patchAmsMathEnvironmentForLineno{#1*}}
  \AtBeginDocument{
  \patchBothAmsMathEnvironmentsForLineno{equation}
  \patchBothAmsMathEnvironmentsForLineno{align}
  \patchBothAmsMathEnvironmentsForLineno{flalign}
  \patchBothAmsMathEnvironmentsForLineno{alignat}
  \patchBothAmsMathEnvironmentsForLineno{gather}
  \patchBothAmsMathEnvironmentsForLineno{multline}
}

\usepackage{xcite}
\usepackage{xr}
\makeatletter
\newcommand*{\addFileDependency}[1]{
  \typeout{(#1)}
  \@addtofilelist{#1}
  \IfFileExists{#1}{}{\typeout{No file #1.}}
}
\makeatother

\usepackage{mathtools}

\DeclareCiteCommand{\citet}
  {\usebibmacro{prenote}}
  {%
    \usebibmacro{citeindex}%
    \printnames{labelname}
    \addspace
    \mkbibbrackets{\printfield{labelnumber}}
  }
  {\multicitedelim}
  {\usebibmacro{postnote}}
  
\title{Functional data decomposition reveals unexpectedly strong soil moisture–precipitation coupling over the Great Plains}

\author{%
  Yifu Gao\textsuperscript{1}, 
  Runze Li\textsuperscript{1},
  Efi Foufoula-Georgiou\textsuperscript{1}, 
  Jasper A. Vrugt\textsuperscript{1}\thanks{Corresponding Author: \texttt{\textcolor{blue}{jasper@uci.edu}}} \\
  \\
  \textsuperscript{1}\small Department of Civil and Environmental Engineering, University of California, Irvine, California, USA
}

\date{} 

\begin{document}

\maketitle

\begin{abstract}
\noindent
Soil moisture–precipitation coupling (SMPC) plays a critical role in Earth's water and energy cycles but remains difficult to quantify due to synoptic-scale variability and the complex interplay of land–atmosphere processes. Here, we apply high-dimensional model representation (HDMR) to functionally decompose the structural, correlative, and cooperative contributions of key land–atmosphere variables to precipitation. Benchmark tests confirm that HDMR overcomes limitations of commonly used correlation and regression approaches in isolating direct versus indirect effects. For example, analysis of gross primary productivity using a light-use-efficiency model shows that linear regression underestimates the temperature effect, while HDMR captures it accurately. Applying HDMR to CONUS404 reanalysis data reveals that morning soil moisture explains up to 40\% of the variance in summertime afternoon precipitation over the Great Plains, more than double prior estimates. On days with afternoon rainfall (12-hour totals of 4.7-8.2 mm), first-order SM effects can boost precipitation by up to 8 mm under wet conditions, with an additional 3 mm from second-order interactions involving temperature and moisture. By capturing real-world co-variability and higher-order effects, HDMR provides a physically grounded, data-driven framework for diagnosing land–atmosphere coupling. These results underscore the need for more nuanced, interaction-aware data analysis methods in climate modeling and prediction.
\end{abstract}


\section{Introduction}
The interaction between soil moisture (SM) and precipitation is a critical driver of Earth's surface water and energy cycles \cite{seneviratne2010}. This SM-precipitation coupling (SMPC) modulates near-surface processes and exerts control on energy partitioning \cite{findell2011probability}, boundary layer dynamics \cite{ek2004influence,ford2023observation}, and mesoscale circulation \cite{taylor2011frequency}, thereby influencing regional extremes of precipitation \cite{levine2016evaluating,dong2024disentangling} and droughts \cite{bevacqua2024direct}. SMPC occurs across a continuum of spatiotemporal scales, spanning distances from a few to several thousands of kilometers and extending from diurnal cycles to seasonal patterns \cite{duerinck2016observed,liu2022influence}. Moreover, SMPC exhibits substantial regional variability in both its strength and sign, attributed to the sensitivity of evapotranspiration to SM and of atmospheric conditions to latent heat fluxes \cite{guo2006glace}. An in-depth understanding of the coupling between SM and precipitation is essential for accurate weather forecasting and climate modeling, especially in the context of global warming and ongoing land-cover and land-use changes.

\begin{figure*}[!ht]
    \centering
    \includegraphics[width=0.8\textwidth]{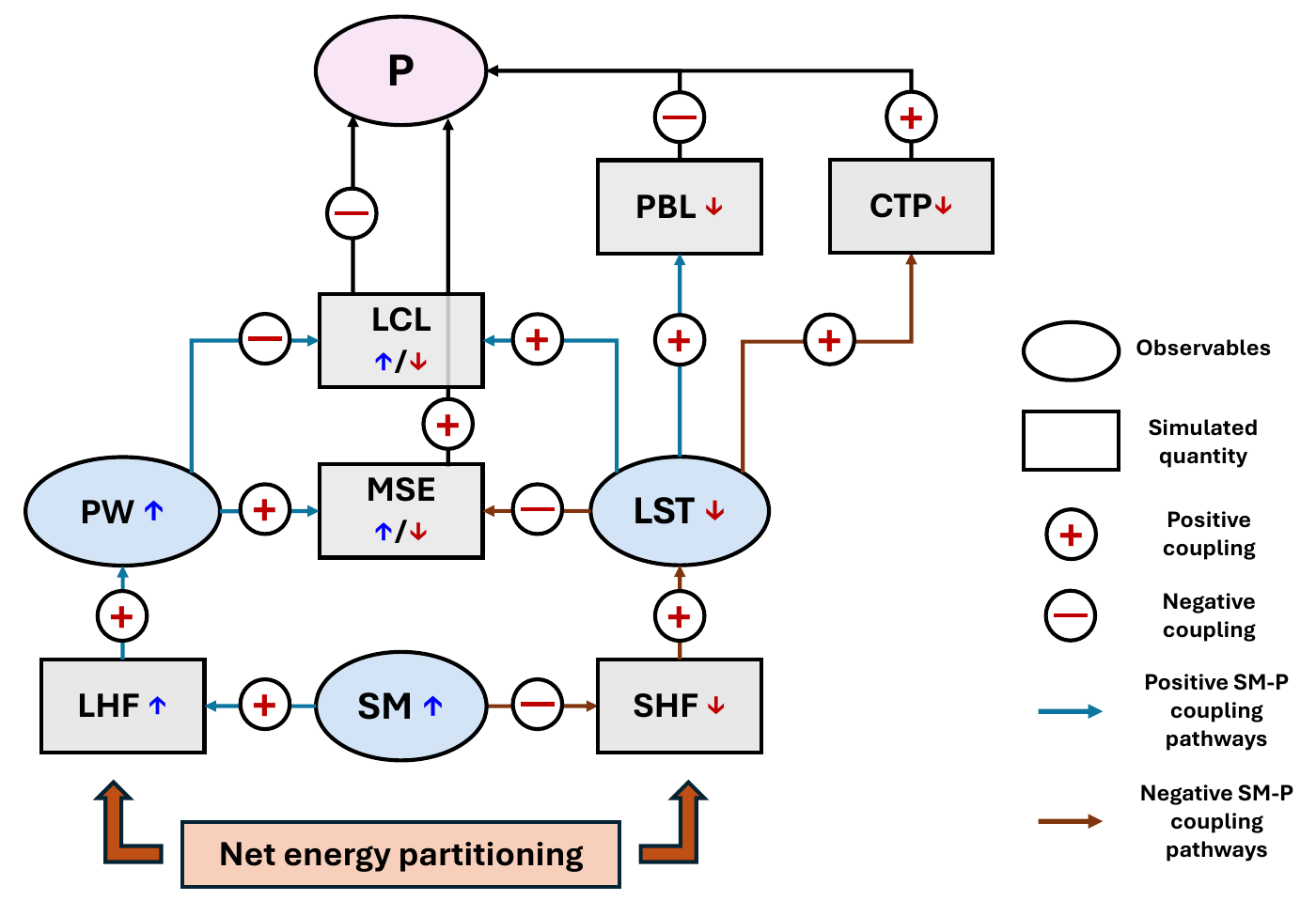}
    \vspace{-4mm}
    \caption{Schematic illustration of positive ("wet-soil effect") and negative ("dry-soil effect") soil moisture–precipitation coupling (SMPC). Key variables include soil moisture (SM), latent heat flux (LHF), sensible heat flux (SHF), land surface temperature (LST), precipitable water (PW), moist static energy (MSE), lifting condensation level (LCL), planetary boundary layer (PBL), and convective triggering potential (CTP). The arrows and interactions demonstrate how wet and dry soil conditions influence boundary-layer moisture, development, and precipitation.}
    \label{fig:1}
\end{figure*}

SMPC mechanisms can have either positive or negative effects on precipitation. These effects are categorized as positive (wet soil) or negative (dry soil) outcomes. As illustrated in Figure \ref{fig:1}, a wet soil supports larger latent heat fluxes (or evapotranspiration) in a SM-limited regime \cite{budyko1974}, increasing the air's moisture content, moist static energy (MSE) \cite{eltahir1998} and evaporative fraction (EF) \cite{findell2011probability}. The triggered convection initiation and shallow planetary boundary layer (PBL) development \cite{taylor2015detecting} promote convective cloud formation and precipitation. Drier soils on the contrary have higher surface temperatures, a larger temperature gradient, and convective triggering potential (CTP) \cite{findell2003a}. Enhanced thermal updrafts in turn reduce convection inhibition (CIN) and promote air parcels to reach the lifting condensation level (LCL) to form precipitation \cite{taylor2012afternoon,ford2015synoptic}. At the mesoscale, both of these feedback mechanisms primarily manifest on a diurnal basis, especially during the midday hours of the warmer seasons. Precipitation recycling is a more straightforward SMPC mechanism but has a more prolonged, non-local, and large-scale impact on the atmosphere \cite{duerinck2016observed,cropper2021comparing}. 

Over the past two decades, many data-driven and simulation-based studies have investigated the strength, sign, and governing mechanisms or pathways of SMPC \cite{findell2003a,findell2011probability,taylor2012afternoon,ford2015synoptic,rappin2022land,ford2023observation}. However, there is still ongoing debate about the extent and causal influence of SM on afternoon precipitation, particularly in hotspot regions such as the Central and Southeastern United States and the Sahel \cite{findell2011probability,taylor2012afternoon,guillod2015reconciling}. One major factor influencing SMPC estimates is the resolution and quality of SM data \cite{taylor2013modeling,yuan2020sensitivity}. Another important, yet underexplored, factor is the variability in how coupling strength is defined and computed across different studies. Most previous work quantifies SMPC using simple statistical metrics or model diagnostics, each with its own set of limitations, as discussed in earlier studies
\cite{seneviratne2010}. Specifically, linear (Pearson) \cite{welty2018does,ford2023observation} or partial correlation coefficients \cite{hu2021early} have been widely used to quantify the direct associations between SM and precipitation or convection initiation. These analyses are often preceded by Principal Component Analysis, which is applied to isolate the leading modes of variability in SM, precipitation, and temperature \cite{chen2019impact,dong2024disentangling}. However, correlation analysis fails to detect nonlinear and multivariate relationships among land–atmosphere variables. Although the coefficients of a multiple linear regression model can provide some insights into coupling strength \cite{zhou2022diminishing, wang2024influence}, such models are inherently linear and additive. They do not distinguish between direct and indirect effects, nor can they capture nonlinear interactions or multivariate dependencies among land-atmosphere variables and precipitation. Composite and Bayesian analyses offer some improvement by examining precipitation patterns under varying SM conditions and classifications \cite{findell2011probability, su2014spring}, but these approaches also fall short in disentangling direct effects from cooperative effects of land–atmosphere variables. Causal inference may help dial in on the direct relationship between SM and precipitation occurrence \cite{tuttle2016} but ignores higher-order synergies and interactions within the land–atmosphere system \cite{li2020causal}. 

Global and regional climate models provide valuable causal insights into SMPC. These models can simulate the effects of SM perturbations on precipitation. This sensitivity is commonly used as a proxy for the strength of land–atmosphere coupling \cite{su2014spring,fast2019impact,wei2021coupling,zhou2021soil} or, alternatively, as the fraction of precipitation variance explained by SM, based on comparisons between reference and prescribed SM climatologies \cite{koster2004,koster2006glace}. 
However, such coupling diagnostics or inferred causal relationships are subject to considerable uncertainties, primarily due to the physical parameterizations of convective and boundary-layer processes within climate models \cite{huang2014evaluation,stephens2010dreary,heinze2017large}. Moreover, artificially imposed SM states used in model experiments often fail to capture realistic antecedent conditions (e.g., the memory effects of prior precipitation) thereby reducing the analysis to a one-way feedback that oversimplifies the full complexity of land–atmosphere interactions. Finally, because these diagnostics rely on controlled perturbations that cannot be reproduced in observational settings, direct validation of the modeled causal relationships is infeasible \cite{seneviratne2010}.

State-of-the-art data and sensitivity analysis methods offer new opportunities to analyze high-dimensional multivariable datasets from both models and observations in the search for robust SMPC signals across spatiotemporal scales. Here, we illustrate the application of one such method -- High-Dimensional Model Representation (HDMR)  \cite{Li2010b,li2012general,gao2023} -- to CONUS404 reanalysis data. HDMR is a generalization of Sobol$'$ \cite{Sobol1993} finite multivariable function expansion to dependent variables, and decomposes the multivariate relationships and coupling among land–atmosphere variables into a hierarchical set of first-, second-, and higher-order response/component functions that capture direct and indirect effects of land–atmosphere variables to precipitation. HDMR's ability to parse out the structural, correlative, and cooperative contributions of land-atmosphere variables to precipitation exceeds the capabilities of commonly used correlation and regression methods and offers deeper insights into the multiple different pathways through which SM modulates precipitation in both modeling and observational frameworks. Our case studies demonstrate HDMR’s potential for inferring causative relationships, quantifying the strength and sign of SMPC across CONUS and diagnosing land-atmosphere coupling in weather and climate models.

\section{High Dimensional Model Representation}
\begin{figure}
\centering\includegraphics[width=0.8\linewidth]{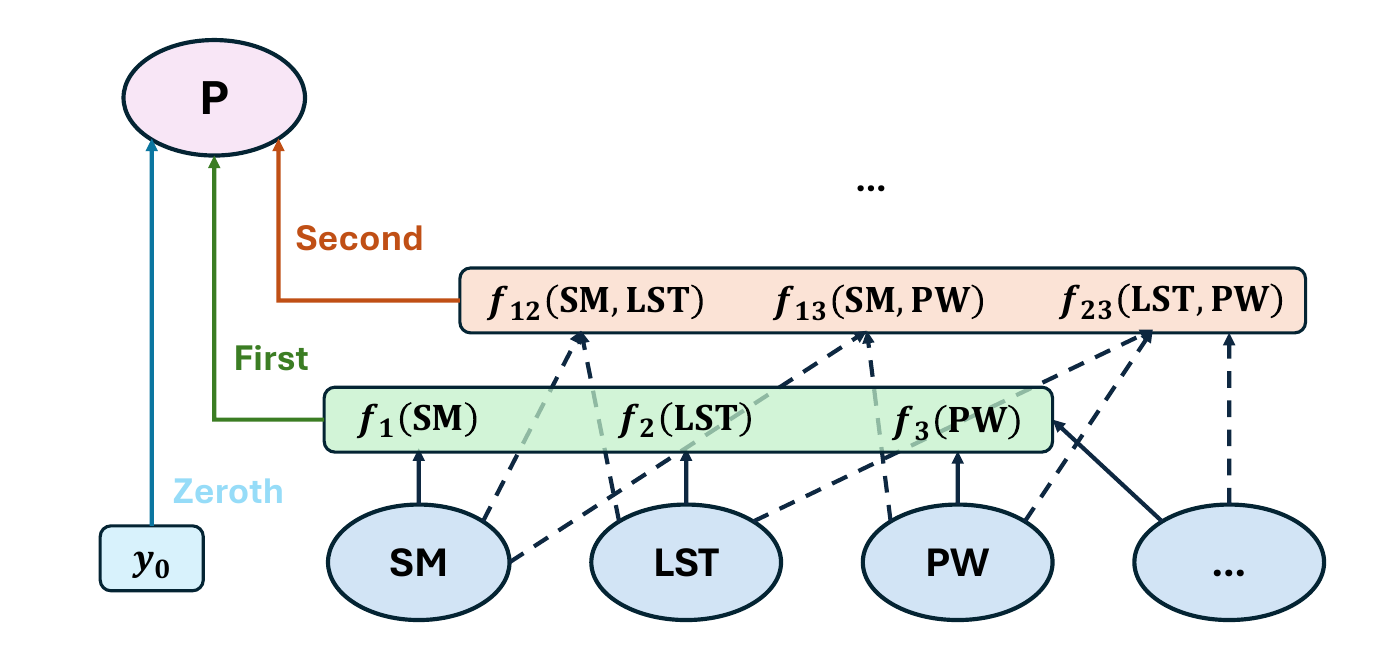}
\vspace{-7mm}
\caption{Schematic overview of the HDMR second-order decomposition of afternoon precipitation as a function of antecedent SM, LST and PW. \vspace{-3mm}}
\label{fig:3}
\end{figure}

Most correlation and regression-based methods capture only direct and linear relationships between SM and precipitation, thereby overlooking the contributions of bivariate and higher-order cooperative effects (e.g., compound contribution of SM and PW) and internal coupling (e.g., SM-LST) of land-atmosphere variables \cite{seneviratne2010}. HDMR, on the contrary, decomposes the SMPC into a hierarchy of component functions and uses an analysis of covariance (ANCOVA) to separate direct (structural) from indirect (correlative and cooperative) effects (see Figure \ref{fig:3}). This enables a more systematic, multivariate and multivariable evaluation of how SM influences precipitation. 

Let $\mathbf{x} = (x_{1},\ldots,x_{d})^{\top}$ be a vector of $d$ input variables (e.g., morning SM and other land-atmosphere variables in Table \ref{table:1}) that explain afternoon precipitation or target variable $y$. HDMR decomposes the afternoon precipitation $y$ into a sum of hierarchical component functions \cite{li2012general}: 
\begin{align}
    y = y_{0} &+ \sum_{i=1}^{n_{1}}f_{i}(x_{i}) + \!\!\!\! \sum^{n_{2}}_{1 \leq i<j\leq d} \!\!\!\! f_{ij}(x_{i},x_{j}) + \!\!\!\!\!\!\! \sum_{1\leq i<j<k \leq d}^{n_{3}} \!\!\!\!\!\! f_{ijk}(x_{i},x_{j},x_{k}) \nonumber\\&+ \dots + f_{12 \dots d}(x_{1},\,x_{2},\dots,x_{d}) + \epsilon,
\label{eq:sobol_decomp_y}
\end{align}
where $y_{0}$ is the mean 12-hr accumulated precipitation between noon and midnight (local time, and on days that experience afternoon rainfall) in units of mm and $\epsilon \sim \mathcal{N}(0,\sigma^{2}_{\epsilon})$ is a zero-mean normally distributed residual with constant variance, $\sigma^{2}_{\epsilon}$. The first-order terms, $f_{i}(x_{i})$, capture the individual contribution of each input variable, while the second- and higher-order terms such as $f_{ij}(x_{i},x_{j}),\dots$, characterize cooperative contributions of two or more variables to $y$. In many physical systems, third- and higher-order terms are often negligible \cite{rabitz1999general,gao2023}. Therefore, we retain only the $n_{12} = n_{1} + n_{2} = d + d(d-1)/2$ first- and second-order component functions:
\begin{equation}
y = y_{0} + \sum\limits_{u=1}^{n_{12}}f_{u}(\textbf{x}_{u}) + \epsilon,
\label{eq:hdmr_model_approx}
\end{equation}
where index $u$ runs over both individual $(d)$ and pairwise $d(d-1)/2$ terms, and $\textbf{x}_{u}$ denotes the subvector corresponding to the individual ($x_{i}$) and pairwise ($x_{i}$,$x_{j}$) components of the input vector $\mathbf{x} = (x_{1},\ldots,x_{d})^{\top}$. The component functions $f_{u}$ quantify the individual and bivariate contributions of the land-atmosphere variables to precipitation. 

To successfully delineate the direct and indirect (cooperative/correlative) effects of the land-atmosphere variables, the component functions must satisfy a so-called \textit{relaxed} vanishing condition \cite{Hooker2007} 
\begin{equation}
\int_{0}^{1} w_{u}(\mathbf{x}_{u})f_{u}(\mathbf{x}_{u})\text{d}x_{i} = 0,
\label{eq:vanishing_condition}
\end{equation}
where $w_{u}(\mathbf{x}_{u})$ is the joint probability density function of the variables in $\textbf{x}_{u}$. \eqref{eq:vanishing_condition} enforces hierarchical orthogonality of the component functions. This condition ensures a unique functional decomposition that properly distinguishes between structural, correlative, and cooperative contributions of the input variables \cite{li2012general,gao2023}. By definition, the relaxed vanishing condition requires each second-order function $f_{ij}(x_{i},x_{j})$ to be orthogonal to its associated first-order counterparts $f_{i}(x_{i})$ and $f_j(x_{j})$. This orthogonality ensures that each second-order HDMR term captures new, interactive information about the target variable $y$ not already represented by lower-order terms. Only then can we successfully unravel the dynamic web of land-atmosphere variables. 

Each component function is represented by a set of orthogonal polynomials $\phi_{r}(x_{i})$ (see Materials and Methods \ref{Sec: Polynomial construction}). The corresponding expansion coefficients are estimated via D-MORPH regression (see Materials and Methods \ref{Sec: D-MORPH regression}) and ensure hierarchical orthogonality. Next, the contribution of each component function $f_{u}(\textbf{x}_{u})$ to the total variance $\Var[y]$ of $y$, is quantified by \emph{structural} and \emph{correlative} sensitivity indices \cite{Li2010a}:
\begin{equation}
S_{u}^{\text{a}} = \frac{\Var[f_{u}(\textbf{x}_{u})]}{\Var[y]} \quad \text{and} \quad S_{u}^{\text{b}} = \frac{\Cov[f_{u}(\textbf{x}_{u}), \sum_{m\neq u}^{n_{12}}f_{m}(\textbf{x}_{m})]}{\Var[y]}. \label{eq:hdmr_struct_corr_indices}
\end{equation}
Their sum defines the coupling index $S_{u}$ of $f_{u}(\textbf{x}_{u})$:
\begin{equation}
S_{u} = S_{u}^{\text{a}} + S_{u}^{\text{b}}.
\label{eq:hdmr_total_index}
\end{equation}
The cooperative coupling index of $x_{i}$ sums up all higher-order coupling indices associated with variable $x_{i}$, $\sum_{j \neq i}S_{ij}(x_{i},x_{j})$. A closely related measure, the total effect \cite{Homma1996}:
\begin{equation}
S_{i}^{\text{T}}(x_{i}) = S_{i}(x_{i}) + \sum_{j \neq i}S_{ij}(x_{i},x_{j}),
\label{eq:sobol_total_effect}
\end{equation}
quantifies the total explained variance of $y$ by variable $x_{i}$ through its direct (structural) and indirect (correlative due to dependencies with other variables and cooperative due to nonlinear interactions) effects. HDMR shares elements with information-theoretic approaches \cite{griffith2014,goodwell2017temporal}, which decompose mutual information into unique, redundant, and synergistic components. This information partitioning, however, assumes  knowledge of the marginal and joint probability density functions (PDFs) of the variables, which can be challenging for high-dimensional datasets. In contrast, HDMR decomposes the target variance into structural, correlative, and cooperative contributions without requiring an underlying PDF, making it computationally more efficient and scalable to large datasets. Below, we present HDMR results for three case studies of increasing complexity. Details of the data, models and methods used in these studies are found in the Materials and Methods.

\section{Results}
\subsection{Case I: A Simple Bivariate Function}
\label{Sec: Results-Case I}
\begin{figure*}[htbp]
    \centering
    \includegraphics[width=0.92\textwidth]{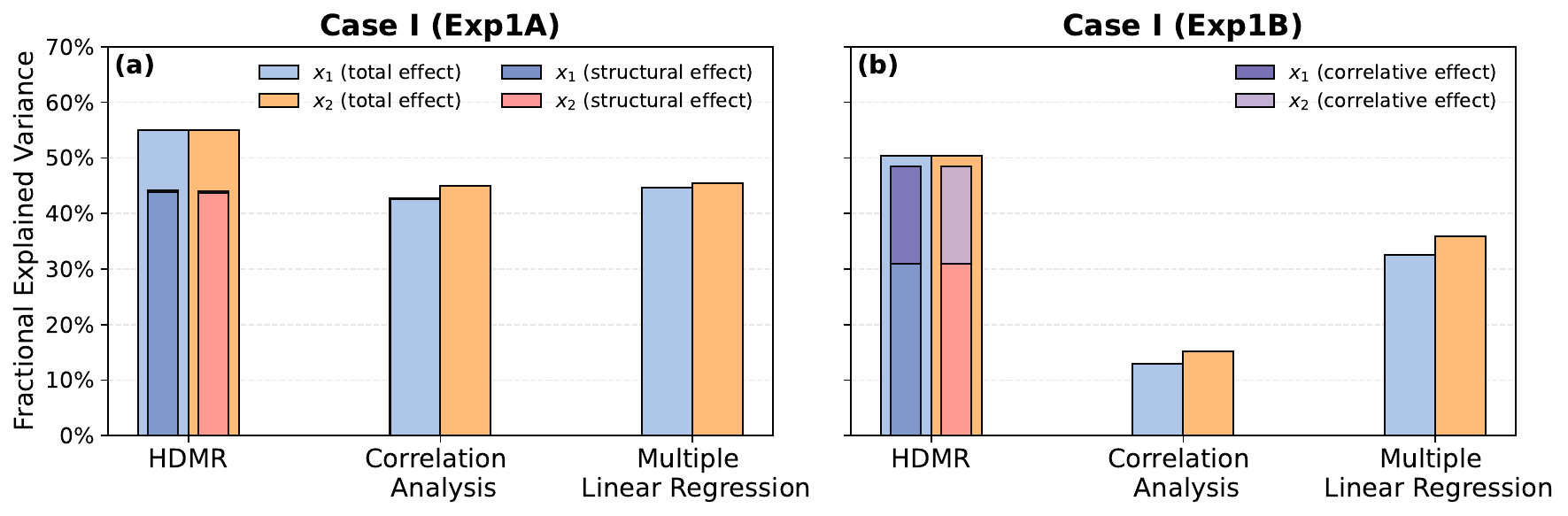}
    \vspace{-3mm} 
    \caption{Fractional contributions of $x_{1}$ and $x_{2}$ to variance of $y$ according to HDMR, Correlation analysis, and Linear Regression for Exp1A (left) and Exp1B (right). The large bars (blue for $x_{1}$ and orange for $x_{2}$) represent each method’s total explained variance. For HDMR, the stacked (smaller) bars show the first-order structural and correlative contributions, respectively. The cooperative (interactive) effect of $x_{1}$ and $x_{2}$ equals the net difference between the HDMR total and the sum of structural and correlative effects.}
    \label{fig:5}
    \vspace{-3mm} 
\end{figure*}

We consider a simple bivariate function $y = x_{1}+x_{2}+x_{1}x_{2}$ and compare the results of HDMR, correlation analysis, and multiple linear regression analysis for two different cases. In Exp1A $x_{1}$ and $x_{2}$ are independent, whereas these two variables are negatively correlated in Exp1B with $r_{x_{1},x_{2}}=-0.6$ (see Materials and Methods \ref{Sec: CASEI Setup} for details on the properties of $x_{1}$ and $x_{2}$). Figure \ref{fig:5} presents the results of our analysis for Exp1A (left) and Exp1B (right) where in each panel, the large bars depict the total fractional variance of $y$ explained by $x_{1}$ (blue) and $x_{2}$ (orange) according to HDMR, correlation analysis, and linear Regression. For HDMR, the smaller stacked bars in dark blue and red portray the first-order structural and correlative contributions, respectively. The smaller purple and pink bars in Exp1B denote the cooperative effect of the second-order component function $f_{12}(x_{1},x_{2})$. 
In Exp1A with uncorrelated inputs, HDMR’s total effects for $x_{1}$ and $x_{2}$ are about 55\% in line with their equal impact on $y$ in \eqref{eq: case 1}. The structural effect of 43.8\% reflects each variable’s individual contribution to $y$. The correlative contribution of 0.3\% for both variables is negligible (not visible in bars) and testifies to the independence of $x_{1}$ and $x_{2}$. The interaction term $S_{12} = S_{1}^{\text{T}} - S_{1}^{\text{a}} - S_{1}^{\text{b}}$ of about 10.9\% accounts for the remainder. The sum of the total effect $\sum S_{i}^{\text{T}}$ does not necessarily equal 1 because the interaction term $S_{12}$ is counted twice in \eqref{eq:sobol_total_effect}. If we compute the normalized total effect $S_{i}^{\text{TN}}(x_{i})$ instead:
\begin{equation}
S_{i}^{\text{TN}}(x_{i}) = S_{i}(x_{i}) + \frac{1}{2}\sum_{i \neq j}S_{ij}(x_{i},x_{j}),
\label{eq:sobol_total_effect_Norm}
\end{equation}
then $S_{1}^{\text{TN}}(x_{1}) = S_{2}^{\text{TN}}(x_{2}) = 43.8 + 11/2 = 49.3$\%. This matches the theoretical expected contribution of 50\% for $x_{1}$ and $x_{2}$ in \eqref{eq: case 1}. Correlation analysis and multiple linear regression report noticeably lower values of 42.6\% and 45.5\% using \eqref{eq: Pearson correlation coefficient_transform}-\eqref{eq: Multiple linear regression_transform}, respectively. These two numbers are in agreement with HDMR's structural effects for $x_{1}$ and $x_{2}$ but omit the interaction altogether. This merely illustrates that correlation and regression-based methods overlook the contribution of the product term $x_{1}x_{2}$ and do not adequately capture cooperative effects on $y$.

The right panel shows that correlation $r_{x_{1},x_{2}} = - 0.6$ among variables $x_{1}$ and $x_{2}$ drastically reduces the fractional explained variance by these two variables according to correlation analysis and multiple linear regression. This correlation does not affect HDMR-based estimates of the total effects of $x_{1}$ and $x_{2}$. Both are locked in at $50$\% and split into a structural part of $30.9$\% and correlative component of $17.5$\%. This time, the interaction term explains only $2$\% of the variance of $y$. This is noticeable less than in the uncorrelated case and this reduction is due to the negative correlation $r_{x_{1},x_{2}} = - 0.6$ of $x_{1}$ and $x_{2}$. These results reiterate HDMR's ability to successfully decompose the variance of $y$ in the presence of variable correlation and interaction. Correlation and regression methods fail in doing so.

\subsection{Case II: A Light-Use-Efficiency (LUE) Model}
\label{Sec: Results-Case II}

\begin{figure*}[htbp]
    \centering
    \includegraphics[width=\textwidth]{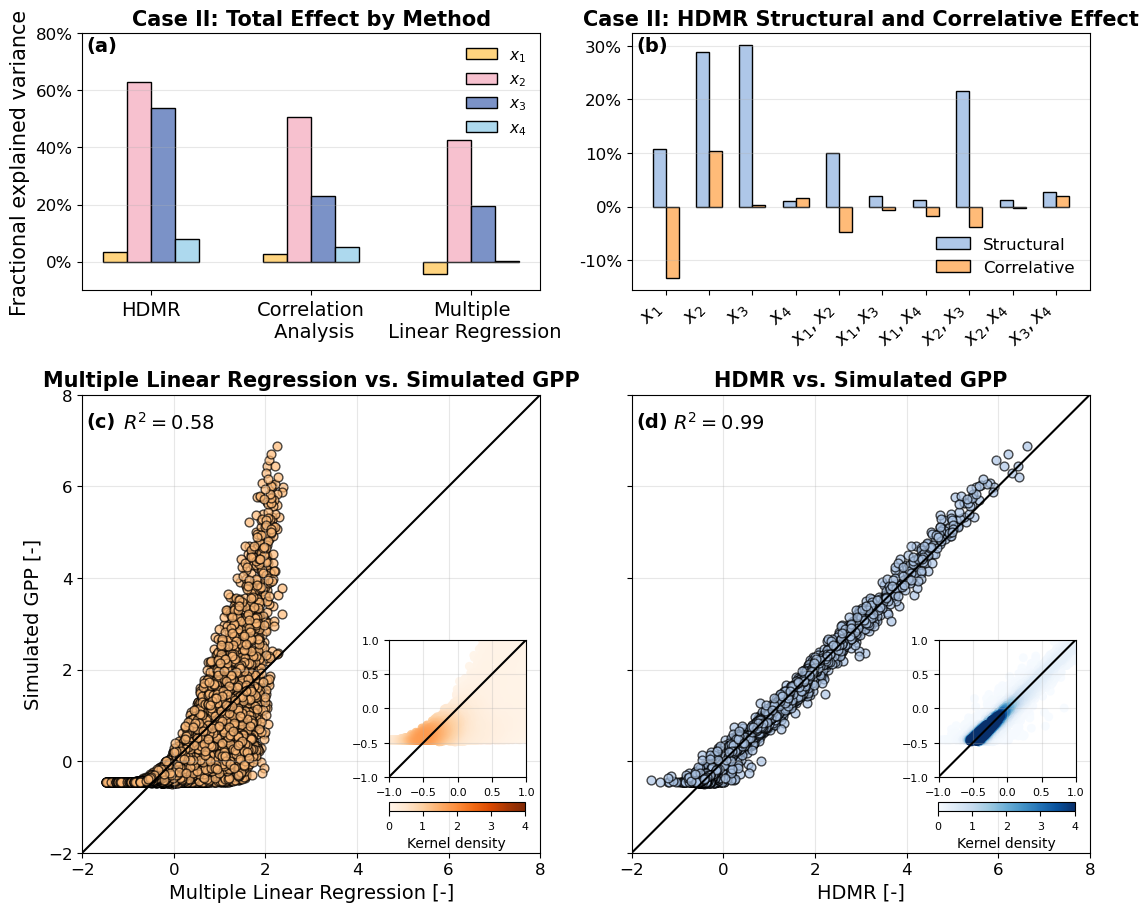}
    \vspace{-3mm} 
    \caption{Top row: (a) total fractional variance of GPP explained by soil moisture $x_{1}$, net radiation $x_{2}$, 2-m air temperature $x_{3}$ and wind speed $x_{4}$ according to HDMR, correlation analysis, and multiple linear regression, and (b) structural and correlative effects of HDMR component functions. Bottom row: scatter plots of (c) multiple linear regression and (d) HDMR predicted GPP values (x-axis) against their simulated values (standardized) from the LUE model, with the 1:1 reference line. The small insets in (c,d) present density estimates of a Gaussian kernel for the [-1,1] data range (Scott bandwidth = 0.22).}
    \label{fig:6}
    \vspace{-3mm} 
\end{figure*}

We use a light use efficiency (LUE) model to simulate data of gross primary productivity (GPP, gC/m$^{2}$/day) as a function of SM, net radiation $R_\text{n}$, 2-m air temperature $T_\text{2m}$, and wind speed $u$. Details are presented in the Materials and Methods \ref{Sec: CASEII Design}. We use this simulated data set to quantify the contributions of $x_{1} = \text{SM}$, $x_{2} = R_\text{n}$, $x_{3} = T_\text{2m}$ and $x_{4} = u$ to GPP. All three methods (HDMR, correlation analysis, and multiple linear regression) are in agreement (see Figure \ref{fig:6}a) on the importance of the four input variables in explaining the variance of GPP. Net radiation exerts the largest control on GPP, followed by 2-m air temperature, SM and wind speed. The effects of SM and wind speed are relatively minimal. Despite assigning similar rankings to the input variables, correlation analysis and multiple linear regression substantially underestimate the explained GPP variance by the different variables. This is particularly true for $T_\text{2m}$ or $x_{3}$. Correlation analysis (22.8\%) and multiple linear regression (19.3\%) substantially underestimate the temperature contribution to GPP. According to HDMR, the explained GPP variance by $T_\text{2m}$ is $50.7$\%. 
To understand these differences, Fig.\ \ref{fig:6}b examines the structural and correlative contributions of the HDMR component functions. The structural effect of $T_\text{2m}$ of about $30$\% is already about $10$\% larger than the total explained GPP variance by correlation analysis and multiple linear regression. This result underscores HDMR’s effectiveness in capturing nonlinear dependencies that are missed by linear methods. In addition, several second-order component functions contribute significantly to GPP, indicating strong interactive effects among the variables. Notably, the joint influence of radiation and temperature $(x_{2}, x_{3})$ explains nearly 20\% of the variance in GPP, an effect that correlation and regression methods fail to capture. The structural and correlative contributions of the component functions jointly explain 99\% of the GPP variance. This inspires confidence in the HDMR functional decomposition of the LUE model. Further evidence for this is presented in Fig. \ref{fig:6}d which shows an excellent agreement ($R^{2}=0.99$) of HDMR predicted and LUE simulated GPP values. Multiple linear regression, on the contrary, explains a much smaller fraction ($R^{2}=0.58$) of the GPP data (see Fig. \ref{fig:6}c).  

\subsection{Case III: Soil Moisture--Precipitation Coupling (SMPC)}
\label{Sec: Results-Case III}

\subsubsection{Hot Spots of SMPC}
We now apply HDMR to CONUS404 reanalysis data to quantify the SMPC strength and identify hotspot regions across the Conterminous United States, focusing on the summer months (April–September) from 2012 to 2021. Data preparation is described in 
sections \ref{Sec: data} and \ref{Sec: CASEIII Design} of the Materials and Method. In short, we examine the 12-hour accumulated precipitation (12:00–24:00 local time), as function of $x_{1}$ morning SM (07:00 local time) and five auxiliary land–atmosphere variables including $x_{2}$ LST, $x_{3}$ PW, $x_{4}$ LAI, and the $x_{5}$ horizontal and $x_{6}$ vertical wind speeds at 10 meter. Synoptic effects were removed by relinquishing samples with antecedant rainfall in the 24 hours leading up to the afternoon events. 

\begin{figure*}[htbp]
    \centering
    \includegraphics[width=0.95\textwidth]{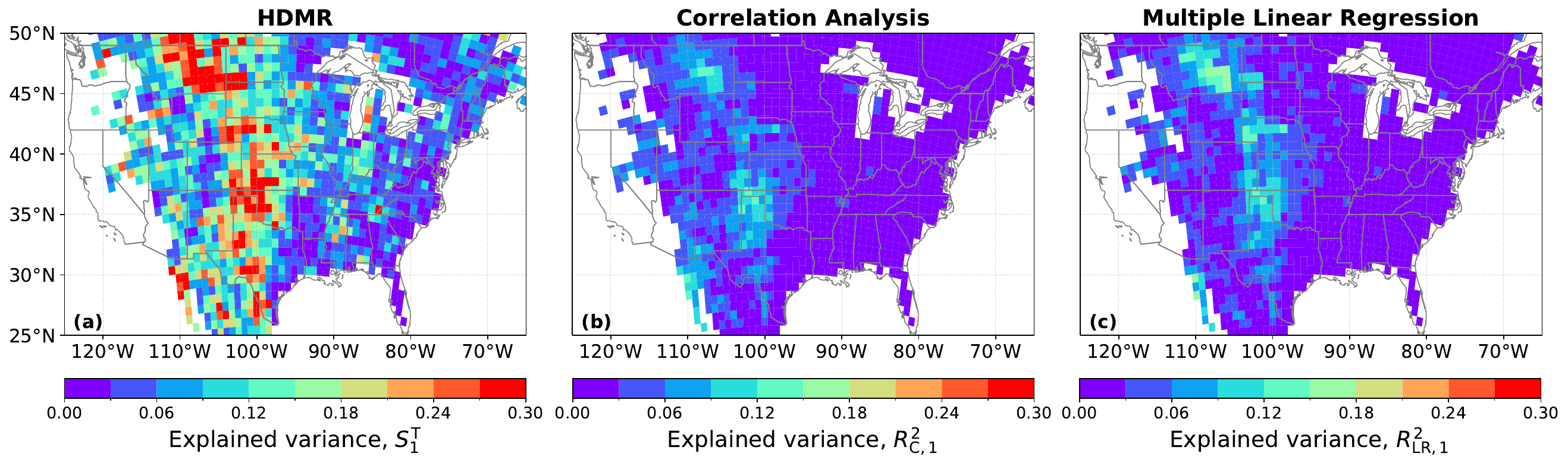}
    \vspace{-3mm} 
    \caption{Total fractional variance of afternoon accumulated precipitation (from noon to midnight) during the warm season (April-September, 2012–2021) explained by morning SM across CONUS: (a) HDMR, (b) correlation analysis, and (c) multiple linear regression. Only grid cells with $5,000$ or more valid data samples are shown in color.}
    \label{fig:7}
    \vspace{-3mm} 
\end{figure*}

Figure \ref{fig:7}a shows HDMR-derived estimates of the total variance in afternoon precipitation over CONUS that is explained by SM alone. We benchmark these results against correlation analysis (Fig. \ref{fig:7}b) and multiple linear regression (\ref{fig:7}c), as these two data analysis methods are commonly used in SMPC studies \cite{duerinck2016observed,welty2018does,zhou2021soil,zhou2022diminishing,zhang2025impacts}.

All three methods identify the Great Plains as a region of relatively strong SM influence on precipitation. The three methods differ substantially in their magnitudes of the explained precipitation variance. According to HDMR, SM explains between 0–40\% of the summer convective precipitation variance over much of the central and northern Great Plains. This is much larger than the 5-15\% derived from correlation analysis and multiple linear regression. Physically, SM plays a particularly influential role in regions where evapotranspiration is water-limited. In these transitional zones, moderate anomalies in soil wetness can have a considerable impact on surface energy fluxes and boundary-layer moisture, thereby influencing convective initiation and rainfall patterns \cite{seneviratne2010}. In contrast, humid regions such as the Southeastern US are characterized by SM abundance and operate under energy-limited conditions, where precipitation is primarily controlled by atmospheric dynamics than by local SM variability or land surface processes \cite{seneviratne2010}. This regime is reflected in the low SM-based explained variance seen in Figure \ref{fig:7}. Nevertheless, SM may still influence the frequency of afternoon rainfall events through a positive feedback between evaporative fraction and precipitation \cite{findell2011probability}. Similarly, the mountainous western US also exhibits relatively weak SM impacts on precipitation, with the notable exception of regions influenced by monsoonal surges (e.g., New Mexico), aligning with previous studies \cite{wang2021impact,gao2024soil}.

Our results are in strong qualitative agreement with GLACE findings \cite{koster2006glace}, but demonstrate a much stronger signal over the Great Plains. According to HDMR, morning SM explains up to 40\% of summertime precipitation variance in this region. This signal is much stronger than the approximately 16\% of Koster et al.\ \cite{koster2006glace}. Two methodological differences likely explain this discrepancy. First, Koster et al.\ used a global climate model with relatively coarse resolution and simplified convection schemes, whereas our analysis is based on CONUS404 -- a high-resolution WRF-derived reanalysis data set that much better captures and describes mesoscale convective systems, boundary-layer dynamics, and precipitation patterns. Second, the GLACE simulations quantify SMPC strength ($\Delta\Omega$) by prescribing soil moisture climatologies. This ``local-sensitivity'' approach decouples the land-atmosphere feedback, suppressing the natural co-variability among land surface and atmospheric variables. In contrast, HDMR is a variance-based global sensitivity analysis method that allows all land-atmosphere variables to vary simultaneously. It partitions precipitation variance into direct (unique) and indirect (shared) contributions from individual drivers and their interactions. As a result, HDMR provides a more physically consistent and substantially higher estimate of SM control on warm-season precipitation. 

\subsubsection{Structural, Correlative, and Cooperative Coupling Effects}
\begin{figure*}[htbp]
    \centering
    \includegraphics[width=0.92\textwidth]{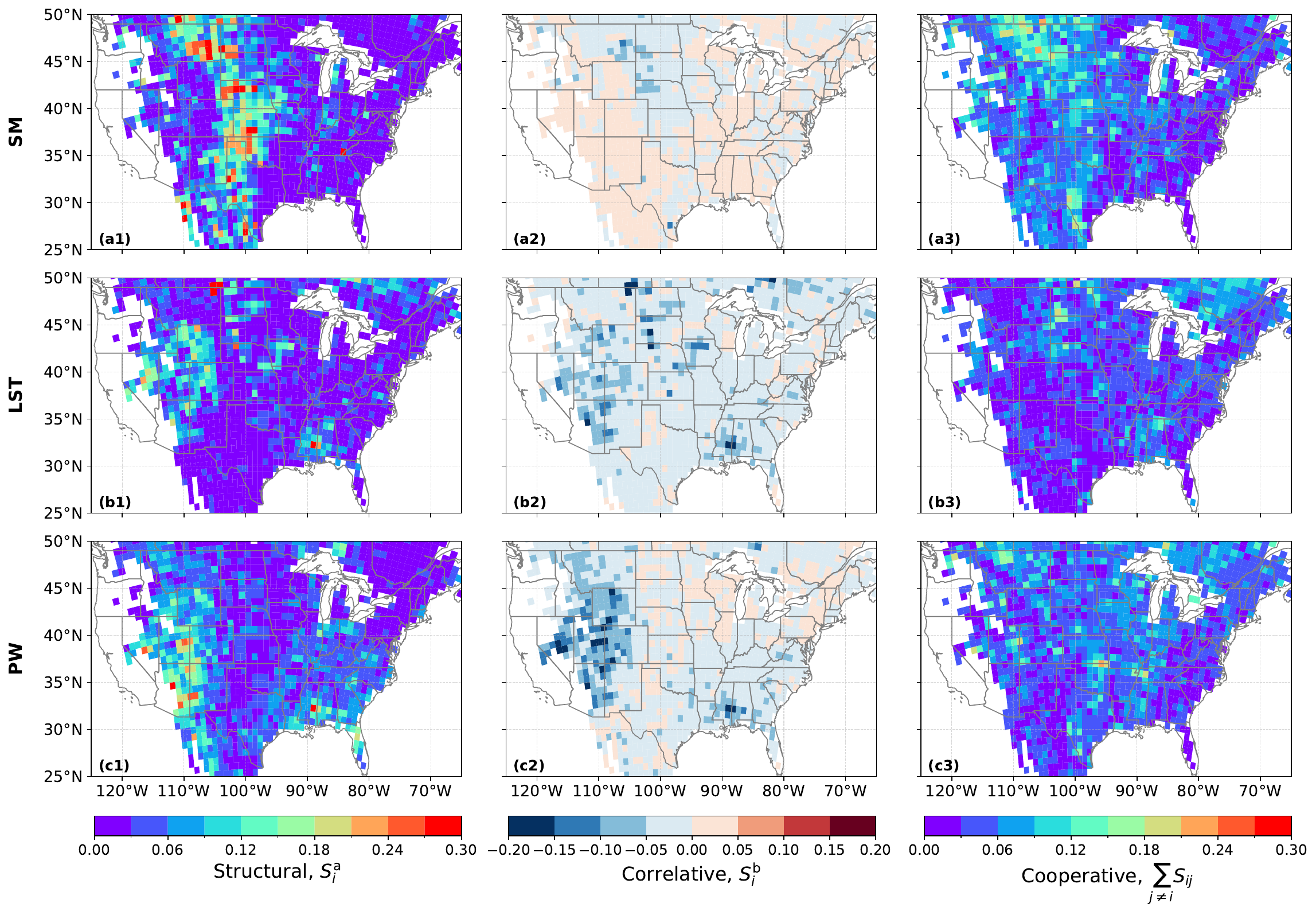}
    \vspace{-4mm} 
    \caption{HDMR results for CONUS: Structural (a1-c1), correlative (a2-c2), and cooperative (a3-c3) coupling indices of (a) SM, (b) LST, and (c) PW.}
    \label{fig:8}
    \vspace{-4mm} 
\end{figure*}

The results in Fig.~\ref{fig:7} raise an important question: Why does HDMR outperform correlation analysis and multiple linear regression in decomposing precipitation variance? Is this a result of model complexity, specifically, the large number of estimable expansion coefficients in the hierarchical polynomial component functions of the HDMR functional decomposition? To investigate this, we retain the exact same polynomial basis but estimate the expansion coefficients using ordinary least squares (OLS) rather than the HDMR-specific D-MORPH regression. The OLS estimator does not enforce the relaxed vanishing condition in \eqref{eq:vanishing_condition}, which is critical for disentangling first- and higher-order structural and correlative effects. For a representative grid cell in the northern Great Plains ($46.2^{\circ}$N, $106.0^{\circ}$W), the OLS-derived SM component function is
\begin{equation}
f_{1}^{\text{OLS}}(x_{1}) = 0.122\phi_{1}(x_{1}) + 0.102\phi_{2}(x_{1}) + 0.068\phi_{3}(x_{1}),
\label{eq: f1_OLS}
\end{equation}
yielding a structural coupling index of $S_{1}^\text{a} = 3$\%. Using the same basis functions, D-MORPH regression produces:
\begin{equation}
f_{1}^{\text{HDMR}}(x_{1}) = 0.433\phi_{1}(x_{1}) + 0.320\phi_{2}(x_{1}) + 0.212\phi_{3}(x_{1}),
\label{eq: f1_HDMR}
\end{equation}
with a substantially higher coupling index of $S_{1}^\text{a} = 34$\%. While both functions explain the same proportion of variance in precipitation ($R^{2} = 57$\%), the OLS parameterization of the SM component function obscures the direct SM effect. In contrast, hierarchical orthogonality enforced by D-MORPH regression uniquely partitions the variance into structural, correlative, and cooperative effects (see Figure~\ref{fig:8}), making the physical role of SM explicit. These findings highlight that HDMR’s superiority lies not in a richer basis, but in its ability to delineate compound influences. This is a prerequisite for mechanistic interpretation of land-atmosphere coupling. 
To further unravel how SM, LST, and PW individually and jointly contribute to precipitation, we analyze their structural, correlative, and cooperative HDMR indices (Figure \ref{fig:8}). LAI, $u_{10}$, and $v_{10}$ have only a minimal contribution to the afternoon precipitation (see Figure S1) and are not discussed further. Provided that SM’s total effect in the Great Plains (Figure \ref{fig:7}, left panel) can exceed 30\%, its structural contribution (Fig.\ \ref{fig:8}a1) makes up almost 80\% of that fraction, indicating a direct and ``stand-alone'' SM impact on precipitation variability. Correlative SM contributions are generally small and possibly even somewhat negative (Fig.\ \ref{fig:8}a2). This finding demonstrates that once the direct contribution of SM to precipitation is accounted for, correlation with other inputs do not systematically amplify precipitation. Yet, for certain hot spots SM displays a moderate cooperative effect (Fig.\ \ref{fig:8}a3) with other variables, underscoring SM's nontrivial interactions with other land-atmosphere variables. Indeed, the nonzero total coupling indices $S_{12}$ and $S_{13}$ in the Northwestern US (see Figure S2) indicate that morning SM, surface heating, and atmospheric humidity jointly influence surface moisture fluxes, CIN, and evaporative cooling, thereby potentially modulating afternoon precipitation.

The roles of LST and PW differ markedly from SM, especially in mountainous and coastal regions. Their structural indices approach 10–25\% over the Rocky mountains (Fig.~\ref{fig:8}b1-c1) and highlights the importance of abundant surface heating and atmospheric moisture for boundary-layer destabilization and moist convection. Yet, some regions exhibit negative correlative indices (Fig.\ \ref{fig:8}b2-c2), where high temperatures coincide with conditions that suppress local rainfall potentially due to dry soils and reduced CAPE \cite{yin2015land}. PW shows a structural effect of up to 25\% in the Southern Rockies and Gulf Coast (Fig.~\ref{fig:8}c1), where moisture-rich air enhances convective potential \cite{kunkel2020observed}. However, in areas frequently affected by large-scale moisture advection, negative correlative contributions diminish PW’s net influence. The cooperative effects (Fig.~\ref{fig:8}b3-c3) of LST and PW are generally modest across CONUS, with localized enhancements along the Southeastern coast, the Southern Rockies, and the Northeastern US-regions where compound processes such as surface heating and tropospheric moisture advection more strongly influence afternoon precipitation 
Overall, the total effects of LST and PW (Figure S3) are comparatively smaller and generally below 15\% across CONUS. This reaffirms the dominant control of SM on warm-season precipitation and its characteristic hot spots within transitional climate zones.

\subsubsection{The Functional SMPC}
\begin{figure*}[htbp]
    \centering
    \includegraphics[width=0.92\textwidth]{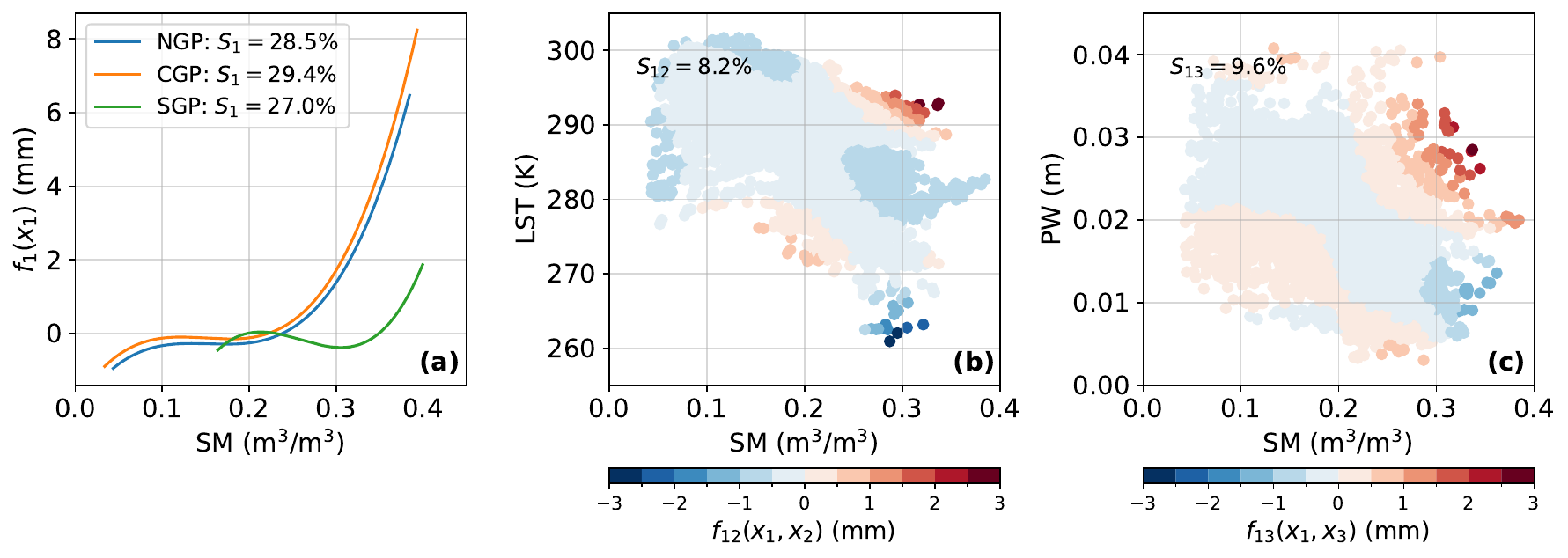}
    \vspace{-3mm} 
    \caption{HDMR results of SMPC in the Great Plains: (a) First-order contribution of SM, $f_{1}(x_{1})$, to afternoon precipitation for three representative grid cells located in the northern ($46.2^{\circ}$N, $106.0^{\circ}$W, blue line), central ($37.7^{\circ}$N, $98.5^{\circ}$W, orange line), and southern ($30.4^{\circ}$N, $100.1^{\circ}$W, green line) Great Plains denoted NGP, CGP, and SGP, respectively, (b,c) Contribution of second-order component functions, $f_{12}(x_{1},x_{2})$ (SM–LST) and $f_{13}(x_{1},x_{3})$ (SM–PW) to precipitation for the NGP grid cell. Coupling indices $S_{1}$, $S_{12}$, and $S_{13}$ quantify the fraction of precipitation variance explained by their respective first- and second-order component functions.}
    \label{fig:9}
    \vspace{-3mm} 
\end{figure*}

The most pronounced SMPC signal is observed over the Great Plains (Fig.\ \ref{fig:7}a), and, thus, we further examine this region to help unlock the physical mechanisms that govern the feedback between SM and precipitation. Fig.~ \ref{fig:9}a displays the first-order SM component functions, $f_{1}(x_{1})$, for three representative grid cells in the northern ($46.2^{\circ}$N, $106.0^{\circ}$W, blue line), central ($37.7^{\circ}$N, $98.5^{\circ}$W, orange line), and southern ($30.4^{\circ}$N, $100.1^{\circ}$W, green line) Great Plains. The mean 12‑hour afternoon rainfall for the analyzed days (i.e., $y_{0}$) at the three grid cells is 4.7, 8.0, and 8.2 mm, respectively. For a cell in the northern Great Plains (NPG), we display the second-order component functions of variable pairs ${\text{SM},\text{LST}}$ and ${\text{SM}, \text{PW}}$ designated $f_{12}(x_{1}, x_{2})$ and $f_{13}(x_{1}, x_{3})$ in Figs. \ref{fig:9}b and \ref{fig:9}c, respectively. The two bivariate interaction terms have relatively high total coupling indices ($S_{12} = 8.2$\% and $S_{13} = 9.6$\%), compared to other SM-related second-order terms (see Figure S2). This warrants a closer inspection.  

In Fig.\ \ref{fig:9}a, the first-order SM component functions exhibit a roughly piecewise relationship between soil wetness and precipitation at the three Great Plains sites, highlighting a distinct ``wet-soil advantage''. 12-hour afternoon precipitation shows minimal sensitivity to soil wetness when SM drops below $0.2$ cm$^{3}$/cm$^{3}$ for the NGP and CGP, and below $0.3$ cm$^{3}$/cm$^{3}$ for the SGP. Above these thresholds, increasing SM leads to enhanced 12-hour precipitation, with peak increases of approximately $6$, $8$, and $2$ mm for the NGP, CGP, and SGP, respectively. These results align closely with the spatial structure of SMPC shown in Fig.\ \ref{fig:7}a1. As expected, the strongest SMPC signals are in the northern and central US, whereas farther south (e.g., in Texas), moisture advection from the Gulf of Mexico and stronger capping inversions likely suppress the local influence of SM on afternoon rainfall \cite{findell2003b}. 


The second-order SM-LST component function $f_{12}(x_{1},x_{2})$ for the NGP (Fig.\ \ref{fig:9}b) indicates  that SM-LST coupling can modulate precipitation by up to $3$ mm when wet soils coincide with high surface temperatures. While the first-order SM component function quantifies the direct contribution of soil water availability to precipitation, the interaction term $f_{12}(x_{1},x_{2})$ reveals how co-occurring high SM and LST jointly enhance evapotranspiration and convective development. Strong solar heating over moist surfaces amplifies moisture fluxes and boundary-layer growth, thereby boosting precipitation. Conversely, low LST suppresses surface fluxes and reduces precipitation, even when SM is abundant. Fig.~\ref{fig:9}c shows that SM-PW interactions can contribute up to 3 mm of additional precipitation when soils are wet ($\text{SM} > 0.30 \; \text{cm}^3/\text{cm}^{3}$) and atmospheric moisture is moderately high ($2 < \text{PW} < 3$ cm). This cooperative effect between SM and PW underscores how moist soils, in tandem with elevated tropospheric humidity, can reduce convective inhibition and promote more efficient rainfall production once storms initiate \cite{cheng2021soil}. Accordingly, while LST and PW alone do not dominate precipitation variability at this NGP location (Fig.\ \ref{fig:8}b,c), their joint interactions with SM explain nearly 10\% of the summertime rainfall variance. Thus, HDMR reveals not only the direct ``wet-soil advantage'' but also key compound effects arising from the co-variability of surface temperature and atmospheric humidity.

\section{Discussion and Conclusions}
\label{sec:discussion}
Soil moisture precipitation coupling (SMPC) exerts strong control on Earth's water and energy cycles, yet is difficult to observe, examine, and quantify due to synoptic effects and the complex and intricate web of land-atmosphere variables involved. Correlation analysis and regression-based methods have become the standard in analyzing SMPCs but are unable to adequately characterize nonlinear and cooperative (compound) effects. In this paper, we introduced functional data decomposition by means of high-dimensional model representation (HDMR) to delineate the structural, correlative, and cooperative contributions of key land-atmosphere variables to precipitation. HDMR \cite{li2012general} is a generalization of Sobol’s functional decomposition \cite{Sobol1993} to dependent (correlated) input variables. Both HDMR and Sobol's method trace their roots to the analysis of variance (ANOVA), which decomposes output variability into contributions from individual inputs and their interactions. While Sobol's approach assumes input independence, HDMR relaxes this assumption, making it suitable for more complex, real-world datasets with correlated drivers. 

A synthetic benchmark experiment with a simple bivariate function showed that HDMR does not suffer the limitations of commonly used correlation and regression analysis methods in distinguishing between direct and indirect (correlative) effects. Data analysis of gross primary productivity (GPP) from a light-use-efficiency model demonstrated that commonly used regression methods substantially underestimate the temperature contribution to GPP, while HDMR accurately captured this effect. Next, continental-scale analysis of CONUS404 reanalysis data confirmed the presence of SMPC hot spots in the central and northern Great Plains. Our regional SMPC characterization aligns closely with earlier modeling studies \cite{koster2004,koster2006glace}, however, the HDMR-based functional decomposition can explain up to 40\% of the variance in summertime afternoon rainfall. This is a substantially higher fraction than the $\sim 16$\% reported both by Koster et al.\ \cite{koster2006glace} and by our linear regression analysis. This difference in the SM coupling strength is attributable to three main factors:
\begin{enumerate}[label=(\roman*)] 
\item We use CONUS404 -- a 4 km WRF-based reanalysis data product -- which explicitly resolves mesoscale convective and boundary-layer processes, preserving the pathways through which SM influences precipitation. \\[-5.5mm]
\item HDMR is a covariance-based global sensitivity analysis method that allows all land–atmosphere variables to vary jointly, instead of varying one at a time. This yields physically consistent estimates of coupling strength. \\[-5.5mm]
\item HDMR enforces hierarchical orthogonality among component functions, allowing it to retain and disentangle structural, correlative, and cooperative effects. Such signals are often missed by simpler data analysis methods.
\end{enumerate}
HDMR results further demonstrate that on summer days with afternoon rainfall (12-hour totals of 4.7 to 8.2 mm), first-order SM effects can increase precipitation by up to 8 mm when soils are sufficiently wet. Second-order effects associated with warm and moist conditions can potentially contribute another 3 mm. 

Distinct roles of land surface temperature (LST) and precipitable water (PW) emerged in mountainous and coastal regions, where surface heating and atmospheric moisture supply usually outweigh local SM effects. While both LST and PW contribute moderately to afternoon precipitation in areas such as the Rocky Mountains and Gulf Coast, their cooperative interactions with SM are particularly variable over the northern Great Plains, highlighting the dynamic nature of land–atmosphere coupling. Collectively, the HDMR results provide a unified framework for understanding how morning SM competes with thermal and moisture conditions in shaping afternoon precipitation. This holistic perspective is particularly valuable in transitional climate zones, where SM variability  can exert substantial control over surface energy partitioning and rainfall intensity.

As with any higher-order polynomial expansion, the HDMR functional decomposition requires substantial data resources. To ensure robust estimates of coupling strength, the number of data samples $N$ must exceed sufficiently the number of expansion coefficients $l$ \cite{gao2023}. This condition will be met with simulated or satellite-based datasets, but HDMR's data requirements increase rapidly with the number of input variables $d$ and order of decomposition, potentially requiring sparse basis representations or regularization techniques. It is also important to note that, although the polynomial component functions in HDMR can capture nonlinear relationships, they are empirical representations of observed (or simulated) input–output behavior rather than explicit mechanistic models. As such, the causal pathways inferred through HDMR must ultimately be validated using process-based modeling and experimental studies.

Our SMPC analysis focuses on a specific diurnal lag, namely, 7:00 am land-atmosphere conditions on same-day afternoon precipitation. However, SMPC is not constrained to a single timescale or lag structure \cite{stacke2016lifetime,wang2018detecting,taylor2024multiday}. Future research can explore wetness-dependent lag dynamics to account for slower, long-memory couplings, particularly under dry conditions where delayed responses may be more pronounced.

We conclude that HDMR offers valuable new insights into SMPC processes and represents a powerful addition to the suite of data analysis tools used by hydrometeorologists and hydroclimatologists. By disentangling the direct, correlative, and cooperative contributions to precipitation variability, HDMR enables a more nuanced diagnosis of land–atmosphere coupling, potentially guiding future improvements in parameterizations of convection and boundary layer processes in weather and climate models.

\newpage
\section{Data, Materials, and Software Availability}

\subsection{CONUS404 Dataset} \label{Sec: data}
The publicly available CONUS404 dataset \cite{rasmussen2023conus404, Rasmussen2023CONUSData}, provides hourly hydroclimate reanalysis at a 4 km spatial resolution over the conterminous United States (CONUS) for water years 1980–2021. This data offers a realistic platform for examining how morning SM and other factors influence afternoon precipitation. We focus on warm-season diurnal SMPC processes by selecting data from April to September over the 2012–2021 period, analyzing afternoon-to-midnight (12:00–24:00 local time) precipitation. Morning conditions of SM and other land–atmosphere variables are used as antecedent states within our SMPC analysis framework. 
\begin{figure*}[htbp]
    \centering
    \includegraphics[width=0.92\textwidth]{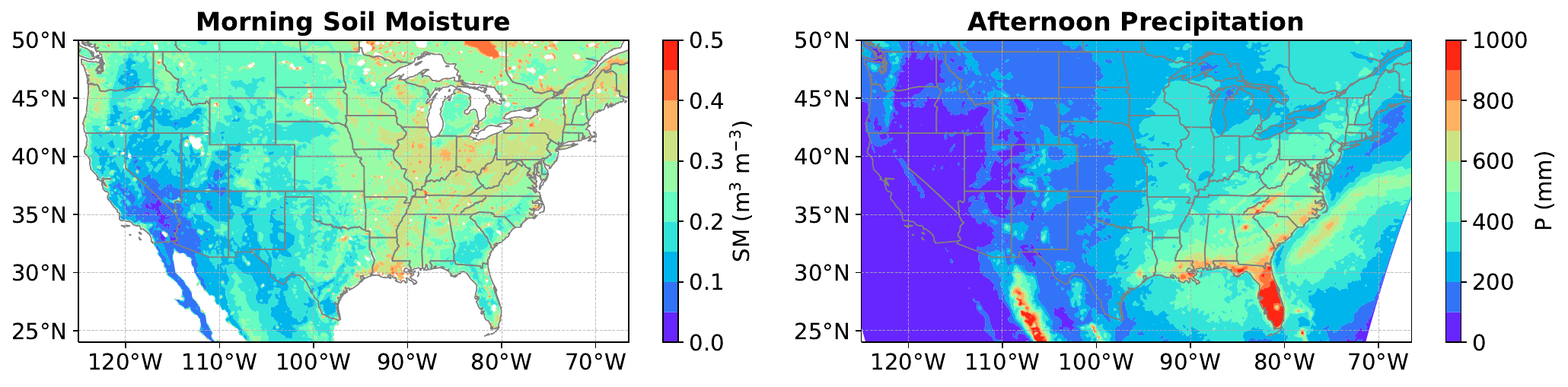}
    \vspace{-3mm} 
    \caption{Ten-year time-averaged values of (a) morning (07:00 local time) soil moisture (SM) and (b) cumulative afternoon precipitation (12:00–24:00 local time) from the CONUS404 reanalysis dataset for the warm seasons (April–September) of 2012–2021. Data are upscaled from 4 km to 16 km using $4 \times 4$ averaging.}
    \label{fig:2}
    \vspace{-3mm} 
\end{figure*}
CONUS404 was generated by dynamically downscaling fifth-generation ECMWF reanalysis (ERA5) using version 3.9.1.1 of the Weather Research and Forecasting (WRF) Model. The primary physical configurations of the CONUS404 reanalysis include the Thompson microphysics scheme \cite{thompson2014study}, Yonsei University planetary boundary layer scheme \cite{hong2006new}, Rapid Radiative Transfer Model for General Circulation Models \cite{iacono2008radiative}, and Noah-MP land surface model \cite{niu2011community}. A key advancement over its predecessor, CONUS1 \cite{liu2017continental}, is the Miguez-Macho–Fan groundwater scheme \cite{miguez2007incorporating,barlage2021importance}, which substantially reduces biases in surface energy fluxes and convective precipitation. Figure~\ref{fig:2} displays the 10-year (2012–2021) mean morning soil moisture (SM) and mean afternoon precipitation (12:00–24:00 local time) during the warm season (April–September).

Table \ref{table:1} list the CONUS404 variables used in this study. Soil moisture (SM), surface skin temperature (LST), precipitable water (PW), leaf area index (LAI), and wind components ($u_{10}$ and $v_{10}$) are used as input variables to HDMR and used to predict afternoon precipitation $P$. These variables characterize key aspects of the land–atmosphere system, including the thermodynamic environment, evapotranspiration, and moisture availability, all of which influence convective triggering and precipitation formation \cite{seneviratne2010, gao2024soil}. Wind speed and direction capture the potential for moisture advection, which can modulate SPMC \cite{ji2024spatially}. Case studies and data preprocessing steps are discussed in the next section. 
\begin{table}[!ht]
\centering
\caption{Land-atmosphere variables of CONUS404 used in this study}
\begin{tabular}{l c c}
\toprule
\multicolumn{1}{l}{ Variables } & Symbol & Units \\
\midrule
Grid-scale cumulative precipitation  & $P$   & mm   \\
Soil moisture                         & SM    & cm$^{3}$/cm$^{3}$   \\
Surface skin temperature              & LST   & K  \\
2-m air temperature                   & $T_{2\text{m}}$   & K \\
Precipitable water                    & PW & m   \\
Leaf area index                       & LAI & -  \\
U-component of wind speed at 10 m     & $u_{10}$ & m/s  \\
V-component of wind speed at 10 m     & $v_{10}$ & m/s  \\
\bottomrule
\label{table:1}
\end{tabular}
\end{table}
\vspace{-6mm}

\subsection{Case Studies and Experimental Design}
We provide the information needed for each case study so that our analyses and numerical experiments can be independently replicated. Specifically, we explain how HDMR's input data were prepared and processed. In each study, HDMR takes as input an $N \times d$ matrix $\mathbf{X}$, consisting of $N$ samples of the input variables $x_{1},\ldots,x_{d}$ and an $N \times 1$ vector $\mathbf{y} = (y^{(1)},\dots,y^{(N)})^{\top}$, containing the corresponding values of the target variable. Prior to executing our HDMR toolbox in MATLAB (or Python), we standardize the entries of both $\mathbf{X}$ and $\mathbf{y}$. As a result, each column of 
$\mathbf{X}$ (i.e., each input variable) has zero mean and unit standard deviation. This preprocessing step promotes consistency across HDMR trials and case studies. 

\subsubsection{Case I: A Simple Bivariate Function}\label{Sec: CASEI Setup}
To verify that HDMR recovers variance partitions under both uncorrelated and correlated inputs, we analyze the scalar-valued function
\begin{equation}
y = x_{1}+x_{2}+x_{1}x_{2}+e.
\label{eq: case 1}
\end{equation}
with $d = 2$ input variables, $x_{1}$ and $x_{2}$, which are sampled from a bivariate normal distribution, $\mathbf{x} \sim \mathcal{N}_{2}(\bm{\upmu},\bm{\Sigma})$ with unit mean $\bm{\upmu} = (1 \; 1)^{\top}$ and $2 \times 2$ covariance matrix, $\bm{\Sigma}$. The noise term, $e \sim \mathcal{N}(0,\sigma_{e}^{2})$ is small compared to $\Var[y]$. We draw at random two data sets, $\mathbf{X}_{1}$ (Exp1A) and $\mathbf{X}_{2}$ (Exp1B), of $N = 5,000$ samples using the covariance matrices:
\begin{equation}
\bm{\Sigma}_{1} = \begin{bmatrix}
1.0 & 0.0\\[1mm]
0.0 & 1.0\\
\end{bmatrix} \quad \quad 
\bm{\Sigma}_{2} = \begin{bmatrix}
1.0 & -0.6\\[1mm]
-0.6 & 1.0\\
\end{bmatrix}.
\label{eq:cov_matrix_case 1}
\end{equation}
The variables $x_{1}$ and $x_{2}$ are independent in data set $\mathbf{X}_{1}$ of Exp1A, whereas they exhibit a negative correlation of $-0.6$ in $\mathbf{X}_{2}$ of Exp1B.

The function output $y$ is linearly dependent on $x_{1}$ and $x_{2}$ individually but also their product (i.e., the interaction term). In Exp1A with uncorrelated inputs, HDMR reduces to Sobol’s method with correlative coupling indices equal to zero. Exp1B demonstrates how correlation among input variables influences the partitioning of structural, correlative, and cooperative effects.

\subsubsection{Case II: Light Use Efficiency Model}\label{Sec: CASEII Design}
The Light Use Efficiency (LUE) model simulates gross primary productivity (GPP) in units of grams of carbon per square meter per day:
\begin{equation}
\mathrm{GPP} = \varepsilon_{0} R_\text{n} \cdot \text{PAR} \cdot f(\text{SM}) \cdot g(T_{2\text{m}}) \cdot h(u), 
\label{eq:GPP}
\end{equation}
where $\varepsilon_{0} = 1.8$ (gC/MJ) is the baseline (maximum) LUE, $R_\text{n}$ (W/m$^{2}$) signifies the net radiation, $\text{PAR} = 0.45$ is the unitless fraction of $R_\text{n}$ that is photosynthetically active radiation and $f(\text{SM})$, $g(T_{2\text{m}})$, and $h(u)$ are dimensionless response functions of SM, 2-meter air temperature and wind speed, respectively. These functions are given below:
\begin{align}
& f(\text{SM}) =
\begin{cases} 
0 & \text{if } \text{SM} < \text{SM}_{\text{w}} \\
\dfrac{\text{SM} - \text{SM}_{\text{w}}}{\text{SM}_{\text{f}} - \text{SM}_{\text{w}}} & \text{if } \text{SM}_{\text{w}} \leq \text{SM} < \text{SM}_{\text{f}} \\
1 & \text{if } \text{SM} \geq \text{SM}_{\text{f}}
\end{cases} \\
& g(T_{2\text{m}}) = \exp[-(T_{2\text{m}} - 25)^{2}/100] \\
& h(u) = \{1 + \exp[-0.5(u - 2)]\}^{-1}, 
\label{eq: GPP NL functions}
\end{align}
where $\text{SM}_{\text{w}} = 0.1$ and $\text{SM}_{\text{f}} = 0.3$ are the soil's wilting point and field capacity in units of cm$^{3}$/cm$^{3}$, respectively.  

We use the CONUS404 reanalysis dataset and extract hourly values of SM, $T_{2\text{m}}$, $u_{10}$, and $v_{10}$ for a grid cell located at $38.80^{\circ}$N$, 97.15^{\circ}$W,  in the central US. Next, we compute the wind speed $u = (u_{10}^2 + v_{10}^2)^{1/2}$ (m/s) and time series of  $R_\text{n}$ (W/m$^{2}$) using Text S1. Then, we evaluate Eq. \eqref{eq:GPP} and simulate a yearly dataset comprised of $N = 8{,}760$ hourly GPP values capturing diurnal to seasonal variability in ecosystem productivity. The standardized LUE model input variables make up the columns of the $N \times 4$ input matrix $\mathbf{X}$, where $x_{1} = \text{SM}$, $x_{2} = R_\text{n}$, $x_{3} = T_{2\text{m}}$ and $x_{4} = u$. The simulated GPP values are standardized and stored in an $N \times 1$ vector $\mathbf{y}$. This $\{\mathbf{X},\mathbf{y}\}$ dataset serves as input to HDMR to estimate the structural, correlative, and interactive contributions of SM, $R_\text{n}$, $T_{2\text{m}}$, and $u$ to plant productivity. 


\subsubsection{Case III: Soil Moisture-Precipitation Coupling (SMPC)}\label{Sec: CASEIII Design}
This study investigates soil moisture–precipitation coupling over the contiguous US. To address this objective, we implement the following CONUS404 data preparation steps. First, to balance computational feasibility and preserve mesoscale variability, the original 4 km CONUS404 outputs are aggregated to 16 km $\times$ 16 km grid cells. Next, we remove data of synoptic systems \cite{findell2011probability,tuttle2017confounding}, as they can obscure local soil moisture impacts. Large-scale stratiform events, for example, can yield a misleading positive correlation between antecedent SM and subsequent rainfall \cite{alfieri2008analysis}. Therefore, we focus attention on precipitation events in the months of April-September of 2012-2021 that satisfy the following two criteria \cite{williams2019evaluating}, (i) the start time, $t_{\text{start}}$, of afternoon rainfall is between 12:00-14:00 local time, (ii) in the antecedant period between $t_{\text{start}}-24$ and $t_{\text{start}}-1$, the grid cell is rain-free. For all grid cells that satisfy the above two criteria, we compute the 12-hr cumulative rainfall between $t_{\text{start}}$ and $t_{\text{start}} + 11$ hr. This rainfall total, expressed in millimeters, constitutes a single entry $y^{(i)}$ in the data vector $\mathbf{y}$. The corresponding morning values at 07:00 local time of SM and the other auxiliary land–atmosphere variables listed in Table \ref{table:1} serve as the explanatory variables for $y^{(i)}$. These morning conditions are stored in the $i$\textsuperscript{th} row of input matrix $\mathbf{X}$, denotes as $\mathbf{x}^{(i)} = (x_{1},\dots,x_{6}) = (\text{SM}, \text{LST}, \text{PW}, \text{LAI}, u_{10}, v_{10})$. 

\begin{figure}
\centering\includegraphics[width=\linewidth]{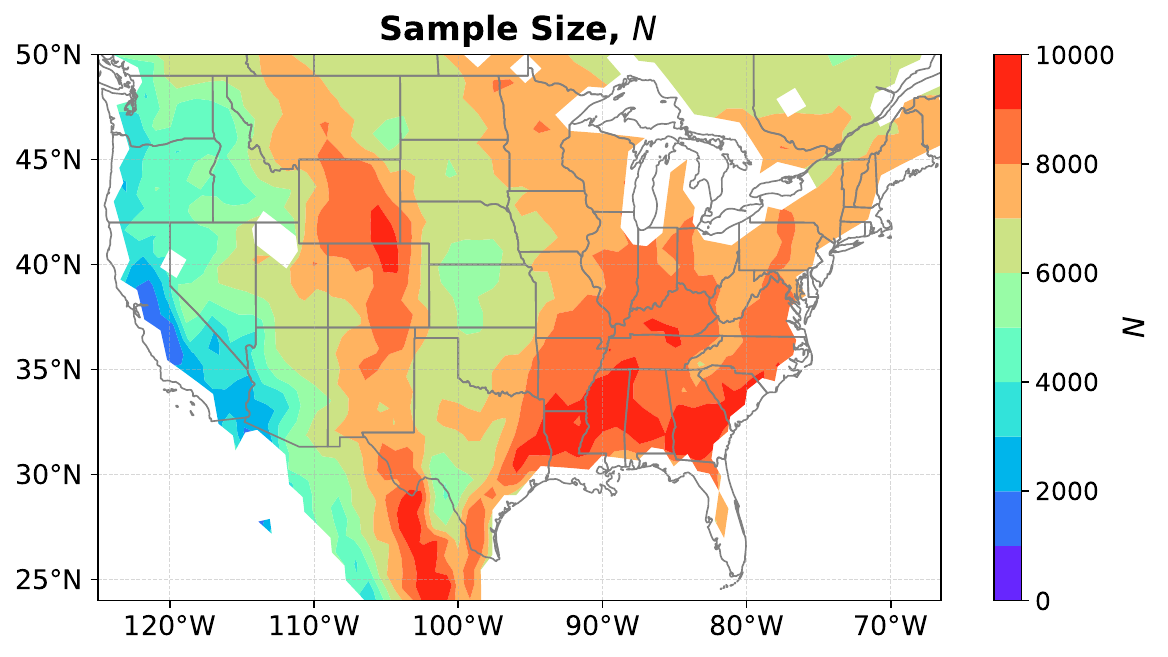}
\vspace{-5mm} 
\caption{Contour map of sample size $N$ used in the HDMR analysis of SMPC, based on a 10-year excerpt (2012–2021) of the CONUS404 dataset for warm-season months (April–September).}
\label{fig:4}
\vspace{-5mm}
\end{figure}

The above data preprocessing steps reduce the number of valid rain-initiation events to only a few hundred for many grid cells across the central US. To satisfy HDMR data requirements, we combine data from $5 \times 5$ blocks of $16 \times 16$ km grid cells in $\mathbf{X}$ and $\mathbf{y}$, corresponding to aggregated areas of $80 \times 80$ km. This upscaling in spatial resolution substantially increases the number of data points $N$, ensuring robust estimation of the polynomial expansion coefficients in the HDMR functional decomposition. As a result, we obtain a single set of coupling metrics for each $80 \times 80$ km block. Figure~\ref{fig:4} displays the sample size $N$ of the HDMR input matrix $\mathbf{X}$ and the $N \times 1$ data vector $\mathbf{y}$ for each aggregated block across CONUS.
Blocks with $N < 5,000$ are discarded from the HDMR analysis (e.g., the southwestern US) as the warm-season precipitation is not frequent enough to yield robust SMPC signals. 

Note that we do not include net radiation ($R_\text{n}$) as an explanatory variable in our HDMR analysis because it is a byproduct of LST and other land–atmosphere variables. Including $R_\text{n}$ would confound the inference of the individual and cooperative effects of LST on afternoon precipitation \cite{gao2024soil}.

\subsection{HDMR Method}
\subsubsection{Polynomial Construction of Component Functions}\label{Sec: Polynomial construction}
We follow Li and Rabitz \cite{li2012general} and construct each component function as a sum of linear multiples of orthonormalized polynomial functions. Specifically, for a single dimension $x_{i}$, we define a family of polynomials:
\begin{subequations}
\begin{align}
    \phi_{1}(x_{i}) & = a_1x_{i} + a_0,\quad & \text{degree }p = 1\hphantom{,} \\
    \phi_{2}(x_{i}) & = b_{2}x_{i}^2+b_1x_1 + b_0, \quad & \text{degree }p = 2\hphantom{,} \\
    \phi_{3}(x_{i}) & = c_{3}x_{i}^3+c_{2}x_{i}^2+c_1x_1 + c_0 \quad & \text{degree }p = 3.
\end{align}
\label{eq:hdmr_ext_polynomial_123}
\end{subequations}
\hspace{-2.3mm} We select $p = 3$ for our case studies. The values of coefficients $a$, $b$ and $c$ are derived from Gram-Schmidt orthonormalization. This process builds an orthonormal basis of polynomial functions on the unit interval with respect to a chosen weighting function. Each HDMR component function then appears as a linear combination of these orthonormalized polynomial functions of degrees $1$ to $p$
\begin{subequations}\label{eq:hdmr_ext_123}
\begin{align}
& f_{i}(x_{i}) = \sum_{r=1}^{p}\alpha_{r}^{(i)i}\phi_{r}(x_{i}) \label{eq:hdmr_ext_1st} \\
& f_{ij}(x_{i},x_{j}) = \sum_{r=1}^{p}\bigl[\alpha_{r}^{(ij)i}\phi_{r}(x_{i}) + \alpha_{r}^{(ij)j}\phi_{r}(x_{j})\bigr] \nonumber \\ 
& \hspace{2cm} \quad + \sum_{r=1}^{p}\sum_{s=1}^{p}\beta_{rs}^{(ij)ij}\phi_{r}(x_{i})\phi_{s}(x_{j}), \label{eq:hdmr_ext_2nd}
\end{align}
\end{subequations}
where the extended bases of the second-order component functions will help satisfy the vanishing condition in \eqref{eq:vanishing_condition}. Parenthesized superscripts in $\alpha$ and $\beta$ indicate which component function (or functions) they belong to, while non-parenthesized superscripts refer to their position in the input vector, $\mathbf{x}$. 

If we substitute \eqref{eq:hdmr_ext_1st} and \eqref{eq:hdmr_ext_2nd} into \eqref{eq:sobol_decomp_y} we obtain a closed-form expression for the relationship between land-atmosphere variables $\mathbf{x} = (x_1, \ldots, x_{d})^{\top}$ and 12-hr accumulated precipitation $y$. This expression has $l = d \cdot p + d(d-1)(2p + p^2)/2$ unknown expansion coefficients and its different terms correspond to the marginal and cooperative effects of the input variables to precipitation. Specifically, the first-order SM component function, $f_{1}(x_{1}) = \sum_{r=1}^{p}\alpha_{r}^{(1)1}\phi_{r}(x_{1})$, quantifies the sign and magnitude (in units of mm) of the SM contribution (direct effect) to precipitation.

\subsubsection{D-MORPH Regression}\label{Sec: D-MORPH regression}
Recall that $\mathbf{x}^{(1)},\dots,\mathbf{x}^{(N)}$ is a $N \times d$ matrix with $N$ different samples of the input variables $x_{1},\ldots,x_{d}$ and $y^{(1)},\dots,y^{(N)}$ is a $N \times 1$ vector with corresponding values of the target variable, then we can write \eqref{eq:hdmr_model_approx} in matrix form $\bm{\Phi}\mathbf{c} = \mathbf{b}$ and yield
\begin{subequations}\label{eq:hdmr_mat_form}
\begin{align}
\bm{\Phi} = \begin{bmatrix}
\bm{\upphi}(\mathbf{x}^{(1)})^{\top} \\
\vdots \\
\bm{\upphi}(\mathbf{x}^{(N)})^{\top}_{\vphantom{\frac{a}{b}}}
\end{bmatrix} \quad \quad &\text{and} \quad \quad  
\mathbf{b} = \begin{bmatrix}
\, y^{(1)} - y_{0} \\
\vdots \\
\, y^{(N)} - y_{0\vphantom{\frac{a}{b}}}
\end{bmatrix},
\label{eq:hdmr_ext_design_matrix}
\end{align}
\end{subequations}
where $\bm{\upphi}(\mathbf{x})^{\top}$ is a $1 \times l$ design vector with orthonormalized polynomial functions of \eqref{eq:hdmr_ext_123} (and products thereof) evaluated at their respective entries of $\mathbf{x}$ and arranged in appropriate order, $\mathbf{c}$ is a $l \times 1$ coefficient vector with values of $\alpha$ and $\beta$, and the $N \times 1$ vector $\mathbf{b}$ stores differences between measured/simulated $y^{(i)}$'s and the mean value $y_{0}$ of the training samples, $i = (1,\ldots,N)$.  

To offer some protection against underdetermined problems $N < l$ or a rank-deficient design matrix, we remove duplicate entries of the basis functions (i.e., $\phi_{r}(x_{i})$) of the first- and second-order component functions. This reduced system is easier to solve in practice \cite{li2012general}. First, we determine the least squares values $\widehat{\mathbf{c}}_\text{ls}$ of the expansion coefficients
\begin{equation}
\widehat{\mathbf{c}}_\text{ls} = (\bm{\Phi}^\top \bm{\Phi})^{\dagger}\mathbf{d},
\label{eq:c_least_squares}
\end{equation}
where the $l \times (l - d \cdot p)$ matrix $(\bm{\Phi}^\top \bm{\Phi})^{\dagger}$ is the generalized pseudo inverse of the $l \times l$ Gramian matrix, $\mathbf{G} = \bm{\Phi}^\top \bm{\Phi}$, which satisfies all four Moore-Penrose conditions \cite{Penrose1955,Golub1996} and whose redundant rows (first $d \cdot p$ rows of the first-order basis functions) are removed and $\mathbf{d}$ is the $(l - d \cdot p) \times 1$ vector $\bm{\Phi}^\top \mathbf{b}$ without the first $d \cdot p$ rows. Diffeomorphic Modulation (dm) under Observable Response Preserving Homotopy (D-MORPH) regression \cite{Li2010b} enforces hierarchical orthogonality of the component functions in pursuit of the optimum coefficients 
\begin{equation}
\widehat{\mathbf{c}}_\text{dm} = \mathbf{V}_{l-r}(\mathbf{U}^{\top}_{l-r}\mathbf{V}_{l-r})\mathbf{U}^{\top}_{l-r}\widehat{\mathbf{c}}_\text{ls},
\label{eq:hdmr_ext_c_dmorph}
\end{equation}
where $\mathbf{U}_{l-r}$ and $\mathbf{V}_{l-r}$ equal the last $l-r$ columns of the $l \times l$ matrices $\mathbf{U}$ and $\mathbf{V}$ determined from singular value decomposition $\mathbf{P}\mathbf{B} = \mathbf{U}\bm{\Sigma}\mathbf{V}^{\top}$ of the product of a $l \times l$ projection matrix $\mathbf{P} = \mathbf{I}_{l} - \mathbf{G}$ and $l \times l$ constraint matrix $\mathbf{B}$ of inner products of the orthonormalized polynomials. Matrix $\mathbf{B}$ enforces the relaxed vanishing condition of \eqref{eq:vanishing_condition} \cite{Li2010b}, $\mathbf{I}_{l}$ is the $l \times l$ identity matrix and $r$ signifies the number of nonzero singular values. 

Our HDMR implementation uses bootstrapping to quantify the uncertainty of the inferred component functions and coupling indices \cite{gao2023}. However, the resulting bootstrap confidence intervals are consistently very narrow, and therefore we do not present them in this paper. This is primarily due to the relatively high ratio of the number of samples $N$ in each $16 \times 16$ km grid cell to the number of expansion coefficients $l$. For example, with $d = 6$ explanatory variables, the third-order $p = 3$ polynomial expansion in our HDMR analysis has $l = d \cdot p + d(d-1)(2p + p^2)/2 = 243$ expansion coefficients. As we only use grid cells with $N>5,000$, the ratio of $N$ and $l$ exceeds 20, and the confidence intervals of the coupling indices are negligibly small. 

\subsection{Other Methods}\label{Sec3: Experimental Setup/Other Methods} 
We benchmark the HDMR results against two other commonly used methods. For each column (input variable) of matrix $\mathbf{X}$ we compute Pearson's linear correlation coefficient, $r_{x_{i},y}$, between $x_{i}$ and $y$ \cite{pearson1895}
\begin{align}
& r_{x_{i},y} = \frac{\sum_{j=1}^{N} (x_{i}^{(j)} - m_{x_{i}})(y^{(j)} - m_{y})}
{\sqrt{\sum_{j=1}^{N} (x_{i}^{(j)} - m_{x_{i}})^2} \sqrt{\sum_{j=1}^{N} (y^{(j)} - m_{y})^2}}, 
\label{eq: Pearson correlation coefficient}
\intertext{where $m_{x_{i}}$ is the sample mean of $x_{i}$ and $i = 1,\ldots,d$. Then, we also use multivariate linear regression}
& y = \beta_{0} + \beta_{1} x_{1} + \beta_{2} x_{2} + \dots + \beta_{d} x_{d} + \epsilon,
\label{eq: Multiple linear regression}
\end{align}
to explain the standardized target variable $y$ from the standardized input variables $x_{1},\ldots,x_{d}$ of matrix $\mathbf{X}$. Values of the intercept $\beta_{0}$ and multiplicative coefficients $\beta_{1},\ldots,\beta_{d}$ are determined using ordinary least squares (OLS). The OLS estimator $\widehat{\bm{\upbeta}} = (\widehat{\beta}_{0},\ldots,\widehat{\beta}_{d})^{\top}$ is equal to $\widehat{\bm{\upbeta}} = (\mathbf{X}^{\top}\mathbf{X})^{-1}\mathbf{X}^{\top}\mathbf{y}$.  

To compare the results of the two methods to the coupling indices of HDMR, we must turn the $r_{x_{i},y}$'s and $\beta_{i}$'s into measures of the normalized explained variance:
\begin{align}
& R_{i,\text{C}}^{2} = r_{x_{i},y}^{2} \label{eq: Pearson correlation coefficient_transform}\\
& R_{i,\text{LR}}^{2} = \frac{\beta_{i}^{2}s_{i}^{2}+\beta_{i}s_{i}\sum_{j\neq i}r_{x_{i},x_{j}}\beta_{j}s_{j}}{\Var[y]},
\label{eq: Multiple linear regression_transform}
\end{align}
where $R_{\text{C},i}^{2}$ and $R_{\text{LR},i}^{2}$ are the explained variances of $y$ by $x_{i}$ according to correlation analysis and multivariate linear regression, respectively, $r_{ij}$ is the correlation coefficient of $x_{i}$ and $x_{j}$, and $s_{i}$ and $s_{j}$ denote the sample standard deviations of $x_{i}$ and $x_{j}$, respectively. This unifies the output of correlation analysis and linear regression with the covariance-based coupling indices of HDMR, allowing direct comparisons.

\subsection*{Data, Materials, and Software Availability}
The CONUS404 dataset is available from the National Center for Atmospheric Research (NCAR) Research Data Archive, as described by Rasmussen et al.\ \cite{Rasmussen2023CONUSData}. MATLAB and Python implementations of the HDMR toolbox can be downloaded from \url{https://github.com/jaspervrugt/HDMR_EXT}. Postprocessing software will be archived on Zenodo along with the final version of the CONUS404-derived dataset.

\subsection*{Acknowledgements}
The authors acknowledge support from NASA through the Precipitation Measurement Mission program (grant 80NSSC22K0597) and the Weather and Atmospheric Dynamics program (grant 80NSSC23K1304), as well as support from the National Science Foundation Division of Information and Intelligent Systems through the Expand AI2ES project (grant IIS-2324008). This research was also partially supported by discretionary funds from the Samueli Endowed Chair awarded to E.F.-G. Yifu Gao gratefully acknowledges support from the Henry Samueli Endowed Fellowship, provided by the CEE Department of the Samueli School of Engineering, UC Irvine.

\printbibliography

@article{taylor2024multiday,
  title={Multiday soil moisture persistence and atmospheric predictability resulting from sahelian mesoscale convective systems},
  author={Taylor, C. M. and Klein, C. and Harris, B. L.},
  journal={Geophysical Research Letters},
  volume={51},
  number={20},
  pages={e2024GL109709},
  doi={10.1029/2024GL109709},
  year={2024},
  publisher={Wiley Online Library}
}

@article{stacke2016lifetime,
  title={Lifetime of soil moisture perturbations in a coupled land--atmosphere simulation},
  author={Stacke, T. and Hagemann, S.},
  journal={Earth System Dynamics},
  volume={7},
  number={1},
  pages={1--19},
  doi={10.5194/esd-7-1-2016},
  year={2016},
  publisher={Copernicus Publications G{\"o}ttingen, Germany}
}

@article{wang2018detecting,
  title={Detecting the causal effect of soil moisture on precipitation using convergent cross mapping},
  author={Wang, Y. and Yang, J. and Chen, Y. and De Maeyer, P. and Li, Z. and Duan, W.},
  journal={Scientific reports},
  volume={8},
  number={1},
  pages={12171},
  doi={10.1038/s41598-018-30669-2},
  year={2018},
  publisher={Nature Publishing Group UK London}
}

@article{williams2019evaluating,
  title={{Evaluating soil moisture feedback on convective triggering: Roles of convective and land-model parameterizations}},
  author={Williams, I. N.},
  journal={Journal of Geophysical Research: Atmospheres},
  volume={124},
  number={1},
  pages={317--332},
  doi={10.1029/2018JD029326},
  year={2019},
  publisher={Wiley Online Library}
}

@article{tuttle2016,
   author    =  "S. Tuttle and G. Salvucci",
   title     =  "Empirical evidence of contrasting soil moisture–precipitation feedbacks across the {U}nited {S}tates",
   year      =  "2016",
   journal   =  "Science",
   volume    =  "352",
   pages     =  "825-828"
}

@article{li2020causal,
  title={A causal inference model based on random forests to identify the effect of soil moisture on precipitation},
  author={Li, Lu and Shangguan, Wei and Deng, Yi and Mao, Jiafu and Pan, JinJing and Wei, Nan and Yuan, Hua and Zhang, Shupeng and Zhang, Yonggen and Dai, Yongjiu},
  journal={Journal of Hydrometeorology},
  volume={21},
  number={5},
  pages={1115--1131},
  doi={10.1175/JHM-D-19-0209.1},
  year={2020}
}

@Inbook{griffith2014,
  author="Griffith, Virgil and Koch, Christof",
  editor="Prokopenko, Mikhail",
  title="Quantifying Synergistic Mutual Information",
  bookTitle="Guided Self-Organization: Inception",
  year="2014",
  publisher="Springer Berlin Heidelberg",
  address="Berlin, Heidelberg",
  pages="159--190",
  isbn="978-3-642-53734-9",
  doi="10.1007/978-3-642-53734-9_6",
  url="https://doi.org/10.1007/978-3-642-53734-9_6"
}

@article{pearson1895,
  author  = {Karl Pearson},
  title   = {Note on Regression and Inheritance in the Case of Two Parents},
  journal = {Proceedings of the Royal Society of London},
  volume  = {58},
  pages   = {240-242},
  year    = {1895},
  doi     = {10.1098/rspl.1895.0041},
  url     = {https://doi.org/10.1098/rspl.1895.0041}
}

@article{goodwell2017temporal,
  title={{Temporal information partitioning: Characterizing synergy, uniqueness, and redundancy in interacting environmental variables}},
  author={Goodwell, A. E. and Kumar, P.},
  journal={Water Resources Research},
  volume={53},
  number={7},
  pages={5920--5942},
  doi={10.1002/2016WR020216},
  year={2017},
  publisher={Wiley Online Library}
}

@article{zhang2025impacts,
  title={{Impacts of land surface processes on summer extreme precipitation in Eastern China: Insights from CWRF simulations}},
  author={Zhang, C. and Li, Q. and Liang, X. and Dong, L. and Xie, B. and Li, W. and Sun, C.},
  journal={Atmospheric Research},
  volume={314},
  pages={107783},
  doi={10.1016/j.atmosres.2024.107783},
  year={2025},
  publisher={Elsevier}
}

@article{zotarelli2010step,
  title={{Step by step calculation of the Penman-Monteith Evapotranspiration (FAO-56 Method)}},
  author={Zotarelli, L. and Dukes, M. D. and Romero, C. C. and Migliaccio, K. W. and Morgan, K. T.},
  journal={Institute of Food and Agricultural Sciences. University of Florida},
  volume={8},
  year={2010}
}

@article{cheng2021soil,
  title={Soil moisture control of precipitation reevaporation over a heterogeneous land surface},
  author={Cheng, Y. and Chan, P. W. and Wei, X. and Hu, Z. and Kuang, Z. and McColl, K. A.},
  journal={Journal of the Atmospheric Sciences},
  volume={78},
  number={10},
  pages={3369--3383},
  doi={10.1175/JAS-D-21-0059.1},
  year={2021}
}

@article{yin2015land,
  title={{Land and atmospheric controls on initiation and intensity of moist convection: CAPE dynamics and LCL crossings}},
  author={Yin, J. and Albertson, J. D. and Rigby, J. R. and Porporato, A.},
  journal={Water Resources Research},
  volume={51},
  number={10},
  pages={8476--8493},
  doi={10.1002/2015WR017286},
  year={2015},
  publisher={Wiley Online Library}
}

@article{wang2021impact,
  title={{Impact of soil moisture initializations on WRF-simulated North American monsoon system}},
  author={Wang, Y. and Quiring, S. M.},
  journal={Journal of Geophysical Research: Atmospheres},
  volume={126},
  number={4},
  pages={e2020JD033858},
  doi={10.1029/2020JD033858},
  year={2021},
  publisher={Wiley Online Library}
}

@article{kunkel2020observed,
  title={{Observed climatological relationships of extreme daily precipitation events with precipitable water and vertical velocity in the contiguous United States}},
  author={Kunkel, K. E. and Stevens, S. E. and Stevens, L. E. and Karl, T. R.},
  journal={Geophysical Research Letters},
  volume={47},
  number={12},
  pages={e2019GL086721},
  doi={10.1029/2019GL086721},
  year={2020},
  publisher={Wiley Online Library}
}

@article{ji2024spatially,
  title={{Spatially varying effect of soil moisture-atmosphere feedback on spring streamflow under future warming in China}},
  author={Ji, P. and Yuan, X.},
  journal={Communications Earth \& Environment},
  volume={5},
  number={1},
  pages={518},
  doi={10.1038/s43247-024-01701-3},
  year={2024},
  publisher={Nature Publishing Group UK London}
}

@article{gao2024soil,
  title={Soil moisture-cloud-precipitation feedback in the lower atmosphere from functional decomposition of satellite observations},
  author={Gao, Y. and Guilloteau, C. and Foufoula-Georgiou, E. and Xu, C. and Sun, X. and Vrugt, J. A.},
  journal={Geophysical Research Letters},
  volume={51},
  number={22},
  pages={e2024GL110347},
  doi={10.1029/2024GL110347},
  year={2024},
  publisher={Wiley Online Library}
}

@article{chen2019impact,
  title={{Impact of nonuniform land surface warming on summer anomalous extratropical cyclone activity over East Asia}},
  author={Chen, H. and Zhan, W. and Zhou, B. and Teng, F. and Zhang, J. and Zhou, Y.},
  journal={Journal of Geophysical Research: Atmospheres},
  volume={124},
  number={19},
  pages={10306--10320},
  doi={10.1029/2018JD030165},
  year={2019},
  publisher={Wiley Online Library}
}

@article{dong2024disentangling,
  title={{Disentangling the Complexities of How Underlying Surface Thermal Factors Influence July Precipitation in Eastern China}},
  author={Dong, X. and Chen, H. and Zhou, Y. and Hsu, P. and Zhang, W.},
  journal={Journal of Climate},
  volume={37},
  number={19},
  pages={5105--5129},
  doi={10.1175/JCLI-D-23-0748.1},
  year={2024},
  publisher={American Meteorological Society}
}

@article{barlage2021importance,
  title={{The importance of scale-dependent groundwater processes in land-atmosphere interactions over the central United States}},
  author={Barlage, M. and Chen, F. and Rasmussen, R. and Zhang, Z. and Miguez-Macho, G.},
  journal={Geophysical Research Letters},
  volume={48},
  number={5},
  pages={e2020GL092171},
  doi={10.1029/2020GL092171},
  year={2021},
  publisher={Wiley Online Library}
}

@article{miguez2007incorporating,
  title={{Incorporating water table dynamics in climate modeling: 2. Formulation, validation, and soil moisture simulation}},
  author={Miguez-Macho, G. and Fan, Y. and Weaver, C. P. and Walko, R. and Robock, A.},
  journal={Journal of Geophysical Research: Atmospheres},
  volume={112},
  number={D13},
  doi={10.1029/2006JD008112},
  year={2007},
  publisher={Wiley Online Library}
}

@article{niu2011community,
  title={{The community Noah land surface model with multiparameterization options (Noah-MP): 1. Model description and evaluation with local-scale measurements}},
  author={Niu, G. Y. and Yang, Z. L. and Mitchell, K. E. and Chen, Fei. and Ek, M. B. and Barlage, M. and Kumar, A. and Manning, K. and Niyogi, D. and Rosero, E. and others},
  journal={Journal of Geophysical Research: Atmospheres},
  volume={116},
  number={D12},
  doi={10.1029/2010JD015139},
  year={2011},
  publisher={Wiley Online Library}
}

@article{iacono2008radiative,
  title={{Radiative forcing by long-lived greenhouse gases: Calculations with the AER radiative transfer models}},
  author={Iacono, M. J. and Delamere, J. S. and Mlawer, E. J. and Shephard, M. W. and Clough, S. A. and Collins, W. D.},
  journal={Journal of Geophysical Research: Atmospheres},
  volume={113},
  number={D13},
  doi={10.1029/2008JD009944},
  year={2008},
  publisher={Wiley Online Library}
}

@article{hong2006new,
  title={A new vertical diffusion package with an explicit treatment of entrainment processes},
  author={Hong, S. Y. and Noh, Y. and Dudhia, J.},
  journal={Monthly weather review},
  volume={134},
  number={9},
  pages={2318--2341},
  doi={10.1175/MWR3199.1},
  year={2006},
  publisher={American Meteorological Society}
}

@article{thompson2014study,
  title={A study of aerosol impacts on clouds and precipitation development in a large winter cyclone},
  author={Thompson, G. and Eidhammer, T.},
  journal={Journal of the atmospheric sciences},
  volume={71},
  number={10},
  pages={3636--3658},
  doi={10.1175/JAS-D-13-0305.1},
  year={2014},
  publisher={American Meteorological Society}
}

@misc{Rasmussen2023CONUSData,
  title={{CONUS404: Four-kilometer long-term regional hydroclimate reanalysis over the conterminous United States (ver. 2.0, December 2023): U.S. Geological Survey data release [Dataset]}},
  author={Rasmussen, R. M. and Chen, F. and Liu, C. and Ikeda, K. and Prein, A. and Kim, J. and Schneider, T. and Dai, A. and Gochis, D. and Dugger, A. and Zhang, Y. and Jaye, A. and Dudhia, J. and He, C. and Harrold, M. and Xue, L. and Chen, S. and Newman, A. and Dougherty, E. and Abolafia-Rozenzweig, R. and Lybarger, N. and Viger, R. and Dunne, K. and Rasmussen, K. and Miguez-Macho, G},
  doi={10.5066/P9PHPK4F},
  year={2023b}
}

@article{liu2017continental,
  title={{Continental-scale convection-permitting modeling of the current and future climate of North America}},
  author={Liu, C. and Ikeda, K. and Rasmussen, R. and Barlage, M. and Newman, A. J. and Prein, A. F. and Chen, F. and Chen, L. and Clark, M. and Dai, A. and others},
  journal={Climate Dynamics},
  volume={49},
  pages={71--95},
  doi={10.1007/s00382-016-3327-9},
  year={2017},
  publisher={Springer}
}

@article{rasmussen2023conus404,
  title={{CONUS404: The NCAR--USGS 4-km long-term regional hydroclimate reanalysis over the CONUS}},
  author={Rasmussen, R. M. and Chen, F. and Liu, C. H. and Ikeda, K. and Prein, A. and Kim, J. and Schneider, T. and Dai, A. and Gochis, D. and Dugger, A. and others},
  journal={Bulletin of the American Meteorological Society},
  volume={104},
  number={8},
  pages={E1382--E1408},
  doi={10.1175/BAMS-D-21-0326.1},
  year={2023a},
  publisher={American Meteorological Society}
}

@article{stephens2010dreary,
  title={Dreary state of precipitation in global models},
  author={Stephens, G. L. and L'Ecuyer, T. and Forbes, R. and Gettelmen, A. and Golaz, J. C. and Bodas-Salcedo, A. and Suzuki, K. and Gabriel, P. and Haynes, J.},
  journal={Journal of Geophysical Research: Atmospheres},
  volume={115},
  number={D24},
  doi={10.1029/2010JD014532},
  year={2010},
  publisher={Wiley Online Library}
}

@article{wei2021coupling,
  title={{Coupling between land surface fluxes and lifting condensation level: mechanisms and sensitivity to model physics parameterizations}},
  author={Wei, J. and Zhao, J. and Chen, H. and Liang, X.},
  journal={Journal of Geophysical Research: Atmospheres},
  volume={126},
  number={5},
  pages={e2020JD034313},
  doi={10.1029/2020JD034313},
  year={2021},
  publisher={Wiley Online Library}
}

@article{huang2014evaluation,
  title={{An evaluation of WRF simulations of clouds over the Southern Ocean with A-Train observations}},
  author={Huang, Y. and Siems, S. T. and Manton, M. J. and Thompson, G.},
  journal={Monthly Weather Review},
  volume={142},
  number={2},
  pages={647--667},
  doi={10.1175/MWR-D-13-00128.1},
  year={2014}
}

@article{su2014spring,
  title={{Spring soil moisture-precipitation feedback in the Southern Great Plains: How is it related to large-scale atmospheric conditions?}},
  author={Su, H. and Yang, Z. and Dickinson, R. E. and Wei, J.},
  journal={Geophysical Research Letters},
  volume={41},
  number={4},
  pages={1283--1289},
  doi={10.1002/2013GL058931},
  year={2014},
  publisher={Wiley Online Library}
}

@article{zhou2022diminishing,
  title={{Diminishing seasonality of subtropical water availability in a warmer world dominated by soil moisture--atmosphere feedbacks}},
  author={Zhou, S. and Williams, A. P. and Lintner, B. R. and Findell, K. L. and Keenan, T. F. and Zhang, Y. and Gentine, P.},
  journal={Nature communications},
  volume={13},
  number={1},
  pages={5756},
  doi={10.1038/s41467-022-33473-9},
  year={2022},
  publisher={Nature Publishing Group UK London}
}

@article{levine2016evaluating,
  title={{Evaluating the strength of the land--atmosphere moisture feedback in Earth system models using satellite observations}},
  author={Levine, P. A. and Randerson, J. T. and Swenson, S. C. and Lawrence, D. M.},
  journal={Hydrology and Earth System Sciences},
  volume={20},
  number={12},
  pages={4837--4856},
  doi={10.5194/hess-20-4837-2016},
  year={2016},
  publisher={Copernicus GmbH}
}

@article{bevacqua2024direct,
  title={{Direct and lagged climate change effects intensified the 2022 European drought}},
  author={Bevacqua, E. and Rakovec, O. and Schumacher, D. L. and Kumar, R. and Thober, S. and Samaniego, L. and Seneviratne, S. I. and Zscheischler, J.},
  journal={Nature Geoscience},
  pages={1--8},
  doi={10.1038/s41561-024-01559-2},
  year={2024},
  publisher={Nature Publishing Group UK London}
}

@article{welty2018does,
  title={{Does soil moisture affect warm season precipitation over the southern Great Plains?}},
  author={Welty, J and Zeng, X},
  journal={Geophysical Research Letters},
  volume={45},
  number={15},
  pages={7866--7873},
  doi={10.1029/2018GL078598},
  year={2018},
  publisher={Wiley Online Library}
}

@article{wang2024influence,
  title={{Influence of lower-tropospheric moisture on local soil moisture--precipitation feedback over the US Southern Great Plains}},
  author={Wang, G. and Fu, R. and Zhuang, Y. and Dirmeyer, P. A. and Santanello, J. A. and Wang, G. and Yang, K. and McColl, K.},
  journal={Atmospheric Chemistry and Physics},
  volume={24},
  number={6},
  pages={3857--3868},
  doi={10.5194/acp-24-3857-2024},
  year={2024},
  publisher={Copernicus Publications G{\"o}ttingen, Germany}
}

@article{zhou2021soil,
  title={Soil moisture--atmosphere feedbacks mitigate declining water availability in drylands},
  author={Zhou, S. and Williams, A. P. and Lintner, B. R. and Berg, A. M. and Zhang, Y. and Keenan, T. F. and Cook, B. I. and Hagemann, S. and Seneviratne, S. I. and Gentine, P.},
  journal={Nature Climate Change},
  volume={11},
  number={1},
  pages={38--44},
  doi={10.1038/s41558-020-00945-z},
  year={2021},
  publisher={Nature Publishing Group UK London}
}

@article{hu2021early,
  title={{Early warm-season mesoscale convective systems dominate soil moisture--precipitation feedback for summer rainfall in central United States}},
  author={Hu, H. and Leung, L. R. and Feng, Z.},
  journal={Proceedings of the National Academy of Sciences},
  volume={118},
  number={43},
  pages={e2105260118},
  doi={10.1073/pnas.2105260118},
  year={2021},
  publisher={National Acad Sciences}
}

@article{tuttle2017confounding,
  title={Confounding factors in determining causal soil moisture-precipitation feedback},
  author={Tuttle, S. E. and Salvucci, G. D.},
  journal={Water Resources Research},
  volume={53},
  number={7},
  pages={5531--5544},
  doi={10.1002/2016WR019869},
  year={2017},
  publisher={Wiley Online Library}
}

@article{guo2006glace,
  title={{GLACE: the global land--atmosphere coupling experiment. Part II: analysis}},
  author={Guo, Z. and Dirmeyer, P. A. and Koster, R. D. and Sud, Y. C. and Bonan, G. and Oleson, K. W. and Chan, E. and Verseghy, D. and Cox, P. and Gordon, C. T. and others},
  journal={Journal of Hydrometeorology},
  volume={7},
  number={4},
  pages={611--625},
  doi={10.1175/JHM511.1},
  year={2006},
  publisher={American Meteorological Society}
}

@article{liu2022influence,
  title={The influence of soil moisture on convective activity: a review},
  author={Liu, W. and Zhang, Q. and Li, C. and Xu, L. and Xiao, W.},
  journal={Theoretical and Applied Climatology},
  volume={149},
  number={1-2},
  pages={221--232},
  doi={10.1007/s00704-022-04046-z},
  year={2022},
  publisher={Springer}
}

@article{koster2006glace,
  title={{GLACE: the global land--atmosphere coupling experiment. Part I: overview}},
  author={Koster, R. D. and Sud, Y. C. and Guo, Z. and Dirmeyer, P. A. and Bonan, G. and Oleson, K. W. and Chan, E. and Verseghy, D. and Cox, P. and Davies, H. and others},
  journal={Journal of Hydrometeorology},
  volume={7},
  number={4},
  pages={590--610},
  doi={10.1175/JHM510.1},
  year={2006}
}

@article{taylor2013modeling,
  title={{Modeling soil moisture-precipitation feedback in the Sahel: Importance of spatial scale versus convective parameterization}},
  author={Taylor, C. M. and Birch, C. E. and Parker, D. J. and Dixon, N. and Guichard, F. and Nikulin, G. and Lister, G. M. S.},
  journal={Geophysical Research Letters},
  volume={40},
  number={23},
  pages={6213--6218},
  year={2013},
  doi={10.1002/2013GL058511},
  publisher={Wiley Online Library}
}

@article{rappin2022land,
  title={{Land--atmosphere interactions during GRAINEX: planetary boundary layer evolution in the presence of irrigation}},
  author={Rappin, E. D. and Mahmood, R. and Nair, U. S. and Pielke Sr, R. A.},
  journal={Journal of Hydrometeorology},
  volume={23},
  number={9},
  pages={1401--1417},
  doi={10.1175/JHM-D-21-0160.1},
  year={2022}
}

@article{Li2010a,
  title = {Global sensitivity analysis for systems with independent and/or correlated inputs},
  author = {Li, G. and Rabitz, H. and Yelvington, P. E. and Oluwole, O. O. and Bacon, F. and Kolb, C. E. and Schoendorf, J.},
  journal = {The Journal of Physical Chemistry A},
  volume = {114},
  number = {19},
  pages = {6022-6032},
  year = {2010},
  sortyear = {2010-1},
  doi = {https://doi.org/10.1021/jp9096919},
  publisher={ACS Publications}
}

@book{budyko1974,
title = "Climate and Life",
author = "M.I. Budyko",
publisher = "Academic Press",
series = "International Geophysics",
volume = "18",
year = "1974",
issn = "0074-6142",
}

@article{ek2004influence,
  title={Influence of soil moisture on boundary layer cloud development},
  author={Ek, M. B. and Holtslag, A. A. M.},
  journal={Journal of hydrometeorology},
  volume={5},
  number={1},
  pages={86-99},
  doi={10.1175/1525-7541(2004)005<0086:IOSMOB>2.0.CO;2},
  year={2004}
}

@article{gao2023,
  author = {Gao, Y. and Sahin, A. and Vrugt, J. A.},
  title = {{Probabilistic sensitivity analysis with dependent variables: Covariance-based decomposition of hydrologic models}},
  journal = {Water Resources Research},
  volume = {59},
  number = {4},
  pages = {e2022WR032834},
  doi = {10.1029/2022WR032834},
  url = {https://agupubs.onlinelibrary.wiley.com/doi/abs/10.1029/2022WR032834},
  eprint = {https://agupubs.onlinelibrary.wiley.com/doi/pdf/10.1029/2022WR032834},
  year = {2023}
}

@article{alfieri2008analysis,
  title={{An analysis of the soil moisture feedback on convective and stratiform precipitation}},
  author={Alfieri, L. and Claps, P. and D’Odorico, P. and Laio, F. and Over, T. M.},
  journal={Journal of Hydrometeorology},
  volume={9},
  number={2},
  pages={280--291},
  year={2008},
  doi={10.1175/2007JHM863.1},
  publisher={American Meteorological Society}
}

@article{findell2003b,
  title={{Atmospheric controls on soil moisture--boundary layer interactions. Part II: Feedbacks within the continental United States}},
  author={Findell, K. L. and Eltahir, E. A. B.},
  journal={Journal of Hydrometeorology},
  volume={4},
  number={3},
  pages={570--583},
  doi={10.1175/1525-7541(2003)004<0570:ACOSML>2.0.CO;2},
  year={2003b}
}

@article{duerinck2016observed,
  title={{Observed soil moisture--precipitation feedback in Illinois: A systematic analysis over different scales}},
  author={Duerinck, H. M. and Van der Ent, R. J. and Van de Giesen, N. C. and Schoups, G. and Babovic, V. and Yeh, P. J. F.},
  journal={Journal of Hydrometeorology},
  volume={17},
  number={6},
  pages={1645--1660},
  year={2016},
  doi={10.1175/JHM-D-15-0032.1},
  publisher={American Meteorological Society}
}

@article{ford2015synoptic,
  title={{Synoptic conditions related to soil moisture-atmosphere interactions and unorganized convection in Oklahoma}},
  author={Ford, T. W. and Quiring, S. M. and Frauenfeld, O. W. and Rapp, A. D.},
  journal={Journal of Geophysical Research: Atmospheres},
  volume={120},
  number={22},
  pages={11--519},
  year={2015},
  doi={},
  publisher={Wiley Online Library}
}

@article{taylor2012afternoon,
  title={Afternoon rain more likely over drier soils},
  author={Taylor, C. M. and de Jeu, R. A. M. and Guichard, . and Harris, P. P. and Dorigo, W. A.},
  journal={Nature},
  volume={489},
  number={7416},
  pages={423--426},
  year={2012},
  doi={10.1038/nature11377},
  publisher={Nature Publishing Group UK London}
}

@article{ford2023observation,
  title = {{Observation-Driven Characterization of Soil Moisture-Precipitation Interactions in the Central United States}},
  author={Ford, T. W. and Steiner, J. and Mason, B. and Quiring, S. M.},
  journal = {Journal of Geophysical Research: Atmospheres},
  volume = {128},
  number = {12},
  pages = {e2022JD037934},
  year = {2023},
  doi = {10.1029/2022JD037934},
  publisher={Wiley Online Library}
}

@article{findell2003a,
  author    =  {Findell, K. L. and Eltahir, E. A. B.},
  title     =  {{Atmospheric controls on soil moisture–boundary layer interactions. Part I: Framework development}},
  year      =  {2003a},
  journal   =  {Journal of Hydrometeorology},
  volume    =  {4},
  doi = {10.1175/1525-7541(2003)004<0552:ACOSML>2.0.CO;2},
  pages     =  {552-569}
}

@article{eltahir1998,
  author = {Eltahir, Elfatih A. B.},
  title = {A soil moisture–rainfall feedback Mechanism: 1. Theory and observations},
  journal = {Water Resources Research},
  volume = {34},
  number = {4},
  pages = {765-776},
  doi = {10.1029/97WR03499},
  url = {https://agupubs.onlinelibrary.wiley.com/doi/abs/10.1029/97WR03499},
  eprint = {https://agupubs.onlinelibrary.wiley.com/doi/pdf/10.1029/97WR03499},
  year = {1998}
}

@article{koster2004,
  author = {Randal D. Koster  and Paul A. Dirmeyer  and Zhichang Guo  and Gordon Bonan  and Edmond Chan  and Peter Cox  and C. T. Gordon  and Shinjiro Kanae  and Eva Kowalczyk  and David Lawrence  and Ping Liu  and Cheng-Hsuan Lu  and Sergey Malyshev  and Bryant McAvaney  and Ken Mitchell  and David Mocko  and Taikan Oki  and Keith Oleson  and Andrew Pitman  and Y. C. Sud  and Christopher M. Taylor  and Diana Verseghy  and Ratko Vasic  and Yongkang Xue  and Tomohito Yamada },
  title = {Regions of strong coupling between soil moisture and precipitation},
  journal = {Science},
  volume = {305},
  number = {5687},
  pages = {1138-1140},
  year = {2004},
  doi = {10.1126/science.1100217},
  url = {https://www.science.org/doi/abs/10.1126/science.1100217},
  eprint = {https://www.science.org/doi/pdf/10.1126/science.1100217}
}

@article{seneviratne2010,
  title = {{Investigating soil moisture–climate interactions in a changing climate: A review}},
  journal = {Earth-Science Reviews},
  volume = {99},
  number = {3},
  pages = {125-161},
  year = {2010},
  issn = {0012-8252},
  doi = {10.1016/j.earscirev.2010.02.004},
  url = {https://www.sciencedirect.com/science/article/pii/S0012825210000139},
  author = {Sonia I. Seneviratne and Thierry Corti and Edouard L. Davin and Martin Hirschi and Eric B. Jaeger and Irene Lehner and Boris Orlowsky and Adriaan J. Teuling}
}

@article{fast2019impact,
  title = {The impact of variable land-atmosphere coupling on convective cloud populations observed during the 2016 {HI-SCALE} field campaign},
  author = {Fast, J. D. and Berg, L. K. and Feng, Z. and Mei, F. and Newsom, R. and Sakaguchi, K. and Xiao, H.},
  journal = {Journal of Advances in Modeling Earth Systems},
  volume = {11},
  number = {8},
  pages = {2629--2654},
  doi = {10.1029/2019MS001727},
  year = {2019},
  publisher = {Wiley Online Library}
}

@article{taylor2011frequency,
  title = {Frequency of {Sahelian} storm initiation enhanced over mesoscale soil-moisture patterns},
  author = {Taylor, C. M. and Gounou, A. and Guichard, F. and Harris, P. P. and Ellis, R. J. and Couvreux, F. and De Kauwe, M.},
  journal = {Nature Geoscience},
  volume = {4},
  number = {7},
  pages = {430-433},
  doi = {10.1038/ngeo1173},
  year = {2011},
  publisher = {Nature Publishing Group UK London}
}

@article{heinze2017large,
  author = {Heinze, R. and Dipankar, A. and Henken, C. C. and Moseley, C. and Sourdeval, O. and Tr\"{o}mel, S. and Xie, X. and Adamidis, P. and Ament, F. and Baars, H. and others},
  title = {{Large-eddy simulations over Germany using ICON: A comprehensive evaluation}},
  journal = {Quarterly Journal of the Royal Meteorological Society},
  volume = {143},
  number = {702},
  pages = {69-100},
  doi = {10.1002/qj.2947},
  url = {https://rmets.onlinelibrary.wiley.com/doi/abs/10.1002/qj.2947},
  eprint = {https://rmets.onlinelibrary.wiley.com/doi/pdf/10.1002/qj.2947},
  year = {2017}
}

@article{guillod2015reconciling,
  title = {Reconciling spatial and temporal soil moisture effects on afternoon rainfall},
  author = {Guillod, B. P. and Orlowsky, B. and Miralles, D. G. and Teuling, A. J. and Seneviratne, S. I.},
  journal = {Nature communications},
  volume = {6},
  number = {1},
  pages = {6443},
  doi = {10.1038/ncomms7443},
  year = {2015},
  publisher = {Nature Publishing Group UK London}
}

@article{taylor2015detecting,
  title = {{Detecting soil moisture impacts on convective initiation in Europe}},
  author = {Taylor, C. M.},
  journal = {Geophysical Research Letters},
  volume = {42},
  number = {11},
  pages = {4631--4638},
  doi = {10.1002/2015GL064030},
  year = {2015},
  publisher = {Wiley Online Library}
}

@article{yuan2020sensitivity,
  title = {{A sensitivity study on the response of convection initiation to in situ soil moisture in the central United States}},
  author = {Yuan, S. and Wang, Y. and Quiring, S. M. and Ford, T. W. and Houston, A. L.},
  journal = {Climate Dynamics},
  volume = {54},
  pages = {2013--2028},
  doi = {10.1007/s00382-019-05098-0},
  year = {2020},
  publisher = {Springer}
}

@article{findell2011probability,
  title = {{Probability of afternoon precipitation in eastern United States and Mexico enhanced by high evaporation}},
  author = {Findell, K. L. and Gentine, P. and Lintner, B. R. and Kerr, C.},
  journal = {Nature Geoscience},
  volume = {4},
  number = {7},
  pages = {434--439},
  doi = {10.1038/ngeo1174},
  year = {2011},
  publisher = {Nature Publishing Group UK London}
}

@article{cropper2021comparing,
  title = {Comparing deuterium excess to large-scale precipitation recycling models in the tropics},
  author = {Cropper, S. and Solander, K. and Newman, B. D. and Tuinenburg, O. A. and Staal, A. and Theeuwen, J. J. E. and Xu, C.},
  journal = {npj Climate and Atmospheric Science},
  volume = {4},
  number = {1},
  pages = {60},
  doi = {10.1038/s41612-021-00217-3},
  year = {2021},
  publisher = {Nature Publishing Group UK London}
}

@article{Sobol1993,
  title = {Sensitivity estimates for nonlinear mathematical models},
  author = {Sobol$'$, I. M.},
  journal = {Mathematical Modelling and Computational Experiment},
  volume = {1},
  number = {4},
  pages = {407-414},
  year = {1993},
  publisher = {John Wiley \& Sons}
}

@book{Golub1996,
  title = {Matrix computations (3rd ed.)},
  author = {Golub, G. H. and Van Loan, C. F.},
  year = {1996},
  pages = {257-258},
  publisher = {Baltimore: Johns Hopkins},
  isbn = {978-0-8018-5414-9}
}

@article{Penrose1955,
  title = {A generalized inverse for matrices},
  author = {Penrose, R.},
  journal = {Proceedings of the Cambridge Philosophical Society},
  volume = {51},
  number = {3},
  pages = {406-413},
  year = {1955},
  doi = {10.1017/S0305004100030401},
  publisher = {Cambridge University Press}
}

@article{Homma1996,
  title = {Importance measures in global sensitivity analysis of nonlinear models},
  author = {Homma, T. and Saltelli, A.},
  journal = {Reliability Engineering \& System Safety},
  volume = {52},
  number = {1},
  pages = {1-17},
  year = {1996},
  doi = {10.1016/0951-8320(96)00002-6},
  publisher = {Elsevier}
}

@article{Li2010b,
  title = {{D-MORPH regression: application to modeling with unknown parameters more than observation data}},
  author = {Li, G. and Rabitz, H.},
  journal = {Journal of Mathematical Chemistry},
  volume = {48},
  pages = {1010–1035},
  year = {2010},
  sortyear = {2010-2},
  doi = {10.1007/s10910-010-9722-2},
  publisher = {Springer}
}

@article{li2012general,
  title={General formulation of {HDMR} component functions with independent and correlated variables},
  author={Li, G. and Rabitz, H.},
  journal={Journal of Mathematical Chemistry},
  volume={50},
  number={1},
  pages={99-130},
  year={2012},
  doi={10.1007/s10910-011-9898-0},
  publisher={Springer}
}

@article{Hooker2007,
  title = {{Generalized functional ANOVA diagnostics for high-dimensional functions of dependent variables}},
  author = {Hooker, G.},
  journal = {Journal of Computational and Graphical Statistics},
  number = {3},
  pages = {709-732},
  volume = {16},
  year = {2007},
  doi = {10.1198/106186007X237892},
  publisher = {Taylor \& Francis}
}

@article{rabitz1999general,
  title = {General foundations of high-dimensional model representations},
  author = {Rabitz, H. and Ali{\c{s}}, {\"O}. F.},
  journal = {Journal of Mathematical Chemistry},
  volume = {25},
  number = {2},
  pages = {197-233},
  year = {1999},
  doi = {10.1023/A:1019188517934},
  publisher = {Springer}
}

\newpage
\section*{Supporting Information}
\subsection*{Text S1. Calculation of Net Radiation from CONUS404 Datasets}
Net radiation, $\text{R}_{\text{n}}$ $\left[\text{W}/\text{m}^{2}\right]$, can be derived from five inputs in the CONUS404 dataset: 
the downward shortwave flux, $\text{R}_{\text{s}}^{\text{in}}$ $\left[\text{W}/\text{m}^{2}\right]$,
the downward longwave flux, $\text{R}_{\text{l}}^{\text{in}}$ $\left[\text{W}/\text{m}^{2}\right]$,
the surface emissivity, $\varepsilon_{\text{s}}$ [-],
the land-surface albedo, $\alpha$ [-],
and the land-surface temperature, LST [K]. 
Using the Stefan–Boltzmann constant 
$\sigma = 5.67\times10^{-8}\,\mathrm{W\,m^{-2}\,K^{-4}}$, 
we express the outgoing longwave flux via the standard blackbody emission formula
\begin{equation}
  \text{R}_{\text{l}}^{\text{out}} \;=\; \sigma \varepsilon_{\text{s}}\; \text{LST}^{4}.
\end{equation}
Following Zotarelli et al.\ \cite{zotarelli2010step}, the net radiation then becomes
\begin{equation}
  \text{R}_{\text{n}} =
  \bigl(\text{R}_{\text{s}}^{\text{in}} - \alpha\text{R}_{\text{s}}^{\text{in}}\bigr)
  - \bigl(\text{R}_{\text{l}}^{\text{in}} - \text{R}_{\text{l}}^{\text{out}}\bigr).
\end{equation}
In other words, this formula accounts for the net shortwave term 
$\bigl(\text{R}_{\text{s}}^{\text{in}}$ minus its reflection due to $\alpha$ and the net longwave term $\text{R}_{\text{l}}^{\text{in}}$ minus the outgoing $\text{R}_{\text{l}}^{\text{out}}$, yielding a physically grounded estimate of $\text{R}_{\text{n}}$. We use this approach for the second case study in the main text, where net radiation is one of the meteorological drivers in the simplified Light Use Efficiency (LUE) model.

\renewcommand{\thefigure}{S\arabic{figure}}
\setcounter{figure}{0}

\clearpage
\begin{figure}
\centering
\includegraphics[width=\textwidth]{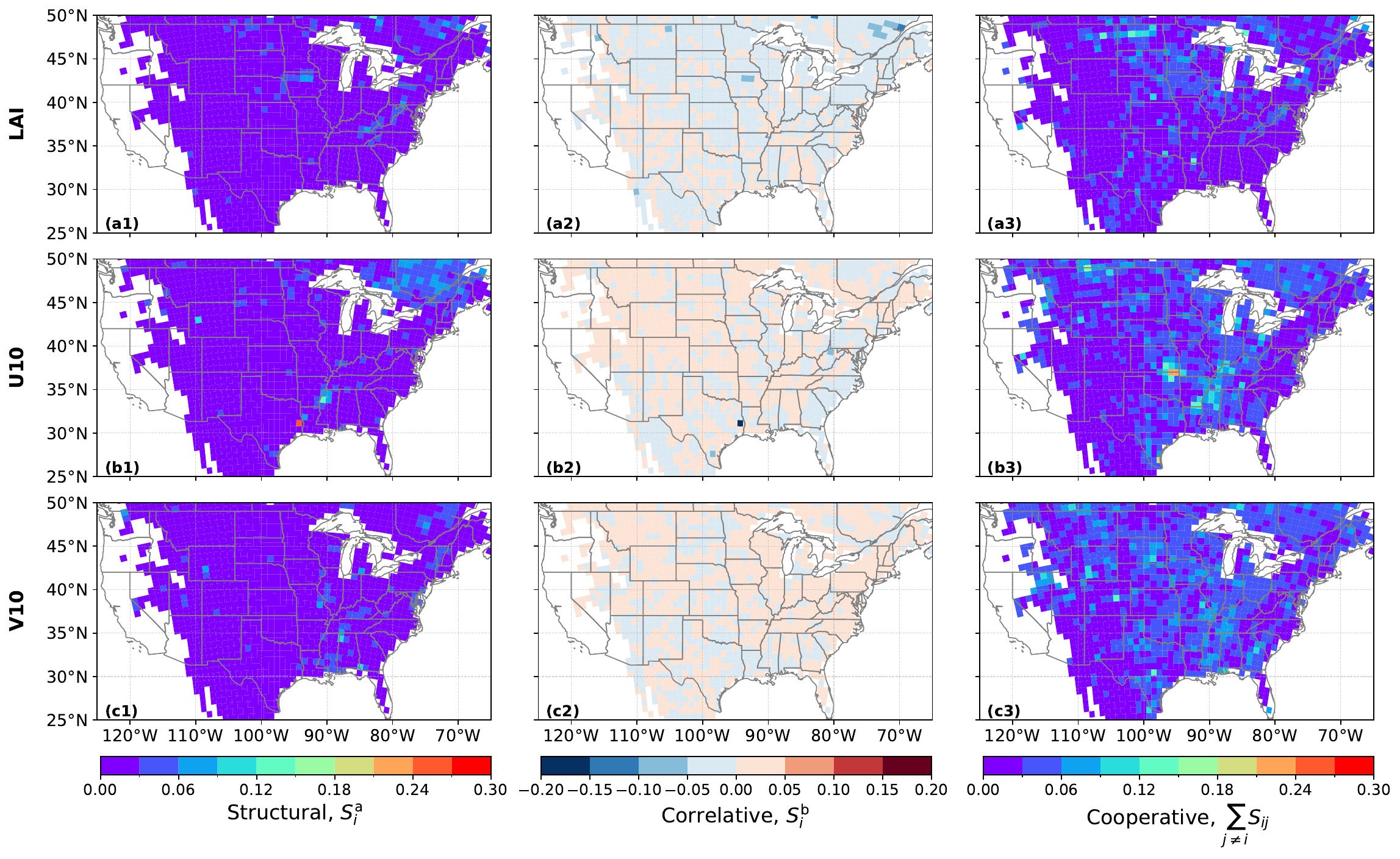}
\caption{Partition of (a) LAI, (b) $\mathbf{u}_{10}$, and (c) $\mathbf{v}_{10}$ into structural, correlative, and cooperative contributions. Each panel (e.g., a1–a3) parallels Figure 8 of the main text, but for variables with minimal overall impact on precipitation. Only grid blocks with at least 5,000 valid rain-initiation events are shown (same for the next two figures).}
\end{figure}

\clearpage
\begin{figure}
\centering
\includegraphics[width=\textwidth]{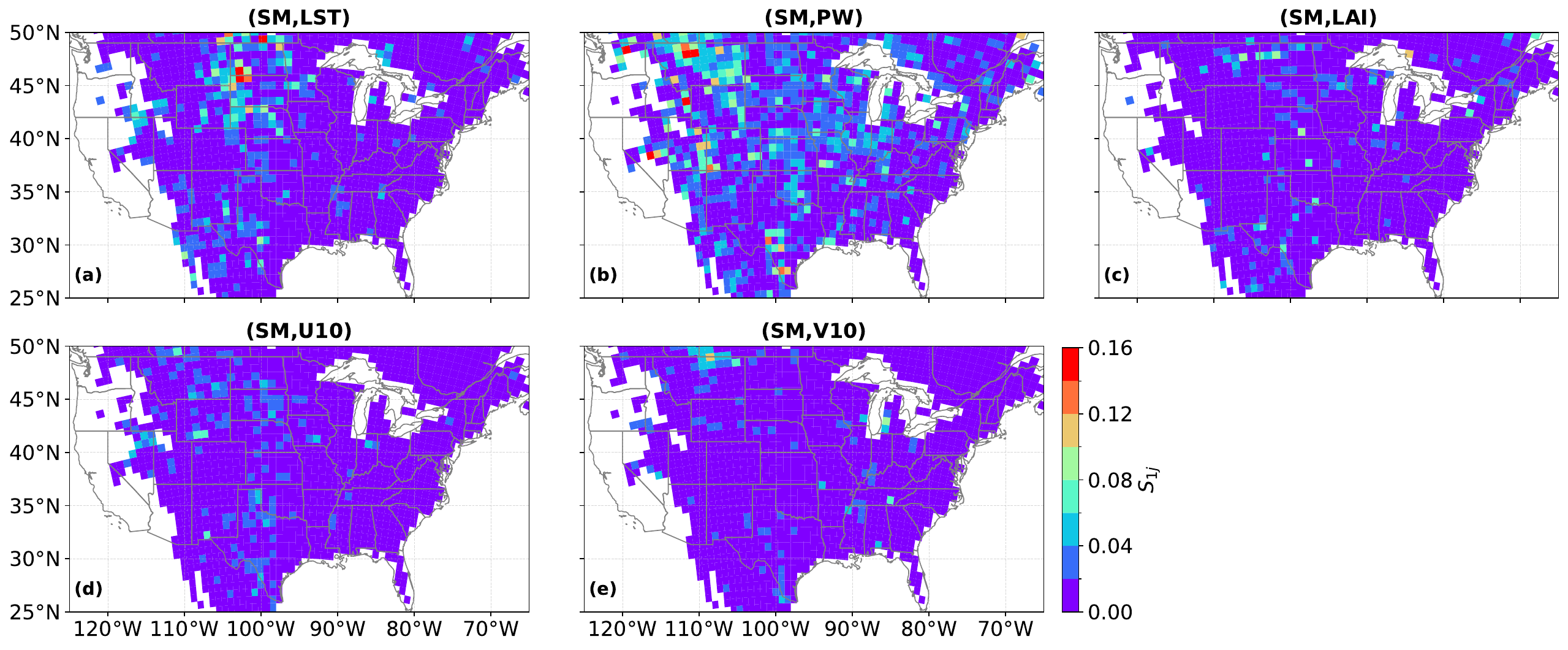}
\caption{Total coupling indices $S_{1j}$ showing cooperative effects between SM and each other variable. Panels (a)–(e) respectively depict the SM–LST, SM–PW, SM–LAI, SM–$\mathbf{u}_{10}$, and SM–$\mathbf{v}_{10}$ cooperative contributions. These maps indicate where morning SM co-varies with surface temperature, atmospheric moisture, vegetation, or wind fields to influence afternoon precipitation beyond their individual (first-order) roles.}
\end{figure}

\clearpage
\begin{figure}
\centering
\includegraphics[width=\textwidth]{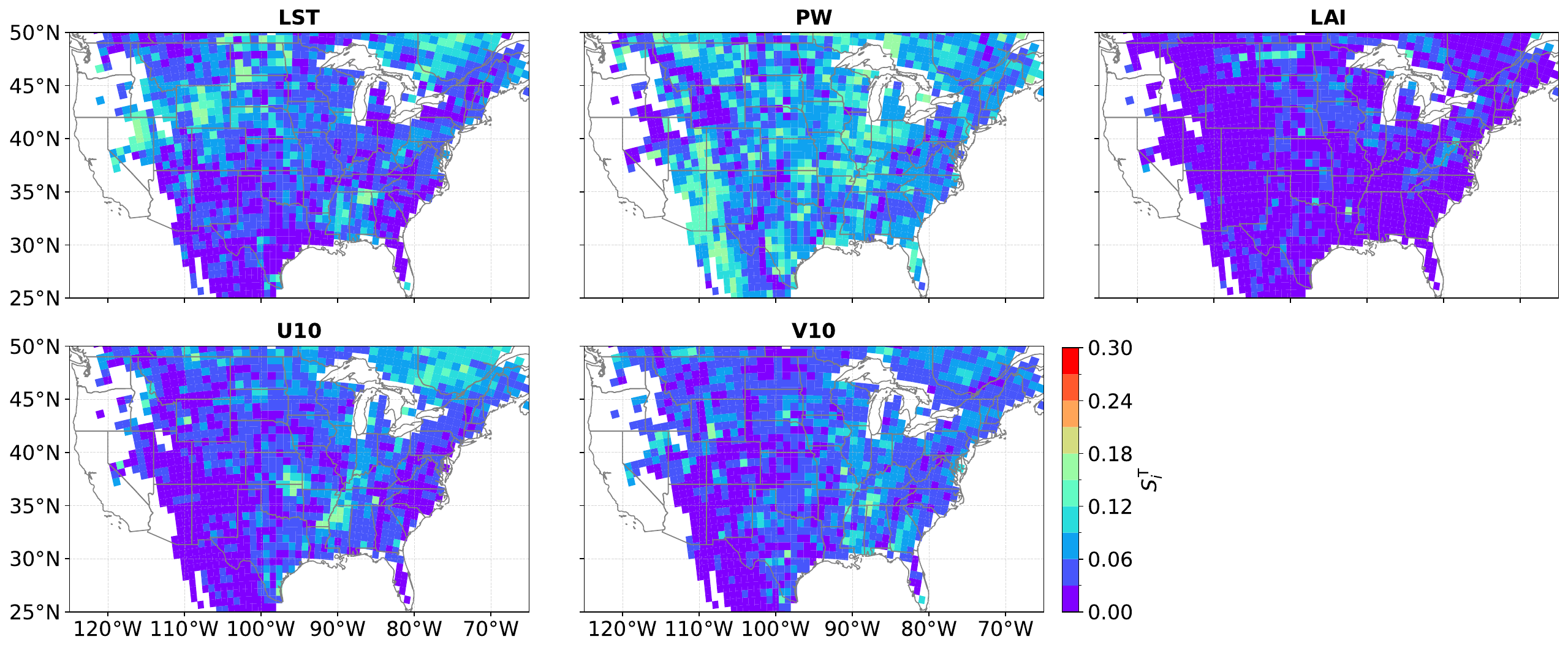}
\caption{Total fractional variance of afternoon precipitation explained by each non-SM variable across the CONUS. Panels (a)–(e) present the HDMR-derived total effect of (a) LST, (b) PW, (c) LAI, (d) $\mathbf{u}_{10}$, and (e) $\mathbf{v}_{10}$, similar to Figure 7 of the main text (which focuses on SM).}
\end{figure}
\end{document}